\documentclass[letterpaper]{article} 
\usepackage{aaai25}  
\usepackage{times}  
\usepackage{helvet}  
\usepackage{courier}  
\usepackage[hyphens]{url}  
\usepackage{graphicx} 
\usepackage{natbib}  
\usepackage{caption} 
\usepackage{algorithm}
\usepackage{algorithmic}

\usepackage{newfloat}
\usepackage{listings}
\DeclareCaptionStyle{ruled}{labelfont=normalfont,labelsep=colon,strut=off} 
\floatstyle{ruled}
\newfloat{listing}{tb}{lst}{}
\floatname{listing}{Listing}
\newcommand{\turbo}[0]{GPT-3.5-turbo}
\newcommand{\xhdr}[1]{\vspace{1.7mm}\noindent{{\bf #1.}}}
\usepackage{amsmath}
\usepackage{booktabs}

\usepackage{xcolor}
\newcommand{\answerYes}[1]{\textcolor{blue}{#1}} 
\newcommand{\answerNo}[1]{\textcolor{teal}{#1}} 
\newcommand{\answerNA}[1]{\textcolor{gray}{#1}}

\title{Capturing Dynamics in Online Public Discourse: A Case Study of Universal Basic Income Discussions on Reddit}
\author {
    Rachel Minyoung Kim\textsuperscript{\rm 1},
    Veniamin Veselovsky\textsuperscript{\rm 2},
    Ashton Anderson\textsuperscript{\rm 3}
}
\affiliations {
    \textsuperscript{\rm 1}Carnegie Mellon University\\
    \textsuperscript{\rm 2}Princeton University\\
    \textsuperscript{\rm 3}University of Toronto\\
    rachelmkim@cmu.edu, veniamin@princeton.edu, ashton@cs.toronto.edu
}

\usepackage{bibentry}

\begin{document}

\maketitle

\begin{abstract}
Societal change is often driven by shifts in public opinion. 
As citizens evolve in their norms, beliefs, and values, public policies change too. 
While traditional opinion polling and surveys can outline the broad strokes of whether public opinion on a particular topic is changing, they usually cannot capture the full multidimensional richness of opinion present in a large heterogeneous population. 
However, an increasing fraction of public discourse about public policy issues is now occurring on online platforms, which presents an opportunity to measure public opinion change at a qualitatively different scale of resolution and context. 

In this paper, we present a conceptual model of observed opinion change on online platforms and apply it to study public discourse on Universal Basic Income (UBI) on Reddit throughout its history. 
UBI is a periodic, no-strings-attached cash payment given to every citizen of a population. We study UBI as it is a clearly-defined policy proposal that has recently experienced a surge of interest through trends like automation and events like the COVID-19 pandemic.
We find that overall stance towards UBI on Reddit significantly declined until mid-2019, when this historical trend suddenly reversed and Reddit became substantially more supportive. Using our model, we find the most significant drivers of this overall stance change were shifts within different user cohorts, within communities that represented similar affluence levels, and within communities that represented similar partisan leanings.
Our method identifies nuanced social drivers of opinion change in the large-scale public discourse that now regularly occurs online, and could be applied to a broad set of other important issues and policies.
\end{abstract}

\section{Introduction}

Societal change is often driven by shifts in public opinion. As people change their minds over time, and as generations change over, attitudes about social norms, justice, and fairness adapt. Public policies are updated to reflect these views, and society evolves. While traditional opinion polling and surveys can outline the broad strokes of whether public opinion on a particular topic is changing, they usually cannot capture the full multi-faceted richness and diversity of opinion present in a large heterogeneous population. 

However, the digital era has ushered in a significant shift in how public discourse is conducted. With the advent of online platforms, a substantial portion of societal conversation, especially regarding public policy issues, has migrated to the digital realm. This transition presents a unique opportunity to observe and measure changes in public opinion with a level of detail and context that was previously unattainable.

Overall trends in public opinion can be driven by qualitatively different mechanisms, with important ramifications depending on which is responsible. The broader ways in which big data systems change over time influence overall online public opinion, which could change as a result of what is being discussed (behavioural drift), who is discussing the subject (population drift), and where the discussion is taking place (system drift)~\cite{salganik2017bit}. Quantifying the mechanisms of macro-level opinion change is important because two contrasting explanations might imply different conclusions. 
For instance, consider a political issue that is typically supported by the left. Stance shift towards the negative on this issue could be driven by an influx of discussion in right-wing communities or a stance change within left-wing communities. While the high-level finding might be the same---stances on a political issue become more negative over time---the implications are quite different. 
The first explanation would highlight that public opinion did not really change between different partisan groups, while the second would highlight intrinsic shift in public opinion within a partisan group.
Understanding the drivers of public opinion changes is critical in ensuring that we are drawing the correct conclusions from our observations.

Previous studies have mostly looked at behavioural~\cite{park_greene_colaresi_2020}, population~\cite{diaz2012online}, and system drifts~\cite{hortaribeiro2023deplatforming} in isolation. 
Comparisons between these different measures and methodologies, however, can offer a more nuanced understanding of macro-shifts. 

In this paper, we present a conceptual model for how macro-level discourse properties can change, which allows us to quantify the impact of various mechanisms of interest. 
We empirically showcase the insights we can get by applying this model on Reddit comments. 
We focus on an important property of discourse, stance, and a real world policy, and measure what drives changes in stance. 
As our public policy of interest, we elect to study Universal Basic Income (UBI)---an unconditional, periodic cash payment provided to every citizen---because it is a clearly-defined policy proposal that is becoming more relevant in mainstream political discourse through trends like automation and events like the COVID-19 pandemic.
We observe that overall stance on UBI experiences a steep decline in June 2016 and continues to become more negative until mid-2019, when the overall stance suddenly becomes more positive.
Then, to rigorously explore a few possible mechanisms that drove the macro-shifts in sentiment, we adopt a novel embedding-based approach of representing communities on Reddit.
The topic-level analysis measures whether stance shifted due to a salience-shift in terms of what UBI sub-issues people care about, or a stance-shift within specific subtopics.
The user-level analysis measures whether stance shifted due to a salience-shift in terms of who discussed UBI, or a stance-shift within specific user cohorts.
The community-level analysis measures whether stance shifted due to a distribution-shift in the communities that discussed UBI (ex. left vs right-wing communities) or a stance-shift within certain communities.
In particular, our key research questions are listed below: 
\begin{itemize}
    \item \textbf{RQ1.} How did stances toward UBI change between 2014--2022?
    \item \textbf{RQ2.} Were changes in stance driven by a distribution shift in which subtopics were discussed most often or a shift in stances towards individual subtopics? 
    \item \textbf{RQ3.} Were changes in stance driven by a distribution shift in the user cohorts that discussed UBI, or a shift in stances within different user cohorts?
    \item \textbf{RQ4.} Were changes in stance driven by a distribution shift in the communities that discussed UBI, or a shift in stances within different communities? 
\end{itemize}

We find stances on UBI change \emph{within} different topics of discussion, users cohorts, and communities of discussion. In each of these cases, the stance shifts are more important than the proportions of different topics, cohorts, and communities that make-up the discussion.
Although intrinsic stance shifts are the most important, proportions for the partisan dimension and proportions across user cohorts still play a statistically significant role in determining overall stance of UBI discussions over time.
Finally, we find that the best explanation for overall stance change is cohort stance change over time.

\section{Related Work}
Our work draws on existing research studying linguistic properties over time.
Computational techniques have enabled the creation of tools like a hedonometer to track Twitter's daily happiness~\cite{dodds2011temporal}.
In works related to political issues, researchers have found that political speeches on immigration have become more positive but increasingly polarized along partisan lines~\cite{card2022immigration}. Furthermore, observing changes in moral language in tweets on same-sex marriage can predict the success of public policy proposals~\cite{zhang2015modeling}.
 
Previous literature has also explored potential drifts in big data sources.
Behavioural drift in topics of discussion has been explored in various domains, from academic conferences~\cite{hall2008studying} to human rights issues~\cite{park_greene_colaresi_2020}. 
Population drift has been explored within the context of political polarization~\cite{waller2021quantifying} and US elections~\cite{diaz2012online}.
System drift caused by community banning and deplatforming have caused behavioural shifts such as migration to other platforms~\cite{hortaribeiro2023deplatforming} and spillover into different communities~\cite{russo2023spillover}, the latter causing increased antisocial behaviour.

Counterfactual models perform “what if?” reasoning by testing alternative hypotheticals. This framework of thinking is often used for causal inference~\cite{hofler2005causal}.Counterfactuals have been used in various domains, including answering causal questions in epidemiology~\cite{hofler2005causal}, assessing the fairness of machine learning models~\cite{kusner2018counterfactual}, and interpreting in language models~\cite{feder2021causalm}. In works most similar to ours, counterfactual thinking was used to study the impact of real-world abortion policies on Twitter discourse~\cite{swanson2023roe}. In our work, we apply counterfactual thinking to the study of online discourse to understand the different factors that drive opinion changes.

UBI is becoming increasingly prominent through trends such as increasing automation and events like the COVID-19 pandemic.
Notably, UBI was the central policy proposed in Andrew Yang's 2020 presidential campaign.
Administrative simplicity~\cite{Nettle2021}, financial anxiety~\cite{Delsen2019}, and women's labour~\cite{robeyns_2001} are themes that are discussed with UBI.  
Today, the policy is favoured by younger~\cite{gilberstadt_2021}, left-wing~\cite{vlandas_2019}, and less affluent~\cite{gilberstadt_2021} people.

In this paper, we illustrate how modern methods let us reconcile the content, population, and behavioural drifts in one methodology to explore public opinion changes surrounding an important policy issue.

\section{First Analysis}

\begin{figure}
    \centering
    \includegraphics[width=7.7cm]{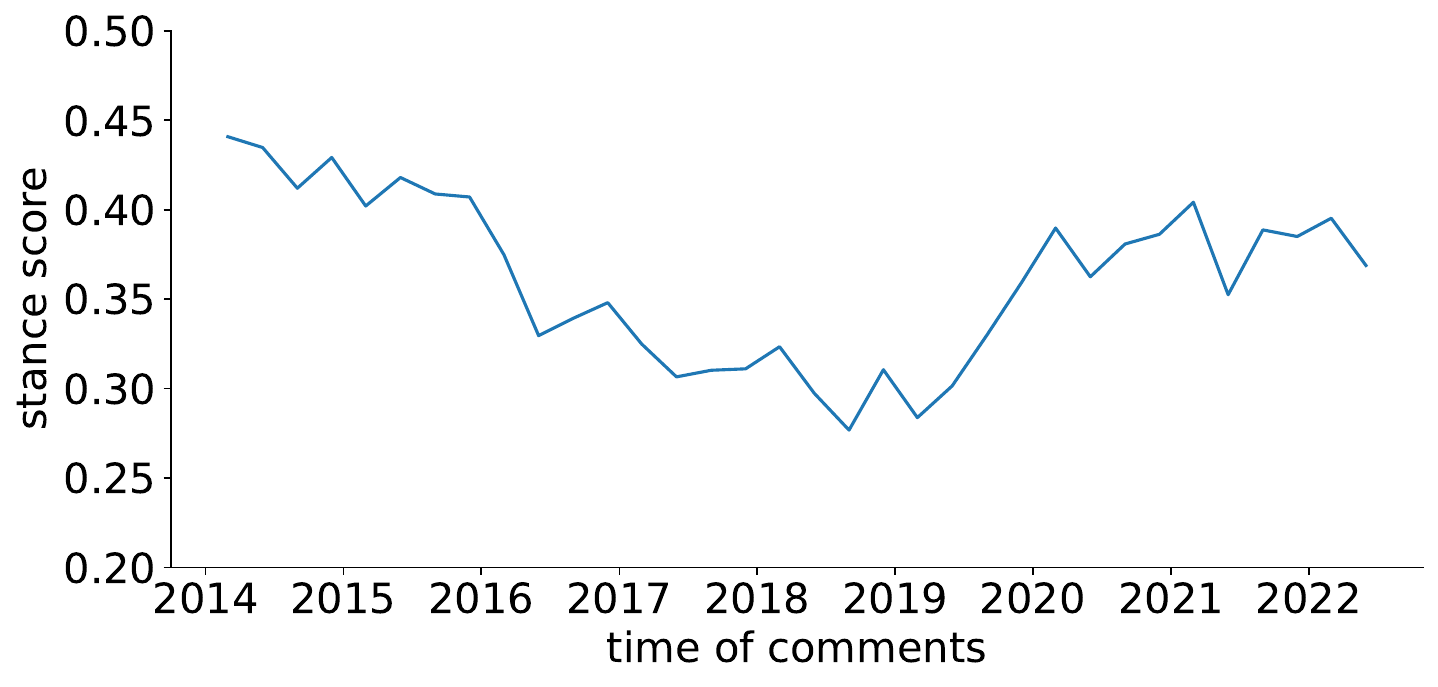}
    \caption{Average stance of UBI comments in each quarter. A comment is assigned a score of $-1$ if it is against UBI, $0$ if it is neutral, and $1$ if it is supportive.}
    \label{fig:overall-stance}
\end{figure}

To motivate our main contributions, we first present a natural ``first attempt'' at a macro-level analysis of how stance surrounding UBI evolved on Reddit from 2014 to 2022. First, assume that we can identify a comprehensive set of comments discussing a topic of interest, e.g., Universal Basic Income (UBI), and furthermore that for each comment we can identify the \emph{stance} of the comment towards UBI (positive, negative, or neutral) with high accuracy. We will demonstrate how to accomplish both of these tasks in the Empirical Methods section. With this information, we can straightforwardly measure the average stance towards UBI platform-wide over time, as in Figure~\ref{fig:overall-stance}. 

This calculation teaches us two important things. First, we learn that supportive comments about UBI on Reddit always outnumber those that are against UBI by a wide margin. Second, the public debate about UBI on Reddit became substantially more negative between June 2016 and March 2019, then it almost completely reversed course and become substantially more positive between April 2019 and June 2022. 
This pattern seems to align with important political events in the U.S.\ such as the polarizing 2016 election and Andrew Yang's UBI-focused Presidential campaign in 2019. 

However, this calculation \emph{doesn't} teach us about perhaps the most important thing: \emph{why} these changes occurred.
The platform-wide distribution of comment stances could have shifted because of changes on the content-level (what is being talked about), user-level (who is talking), or community-level (where discussion is taking place). 
Within each of these potential drivers, there are two distinct ways that each driver could contribute to overall stance change: a shift in the driver's underlying distribution, and a shift in the relationship between driver and stance.
For example, considering content-level drivers, overall stances could become more negative because of the influx of new UBI subtopics that tend to be more negative (for example, ``Money and inflation'' could be a more negative subtopic that starts becoming heavily discussed) or the discussions surrounding existing subtopics becoming more negative.

In order to understand the drivers of online opinion change, we need a method capable of disambiguating between these potential mechanisms.  

\section{Conceptual Model}\label{sec:theory}

How should we think about the relationship between overall opinion and the various mechanisms we have discussed? Here we formalize a conceptual model to understand the extent to which each of these drivers contributes to overall stance changes such as those we observed in Figure~\ref{fig:overall-stance}. 

Consider a social media platform with a set of posts $P = \{p_1, ..., p_n\}$, and a target construct $L$ which is measured through a proxy $\hat{L}$. Each post $p_i$ has a corresponding metadata tuple $(\hat{l}_i, \mathbf{s_i}, \mathbf{u_i}, \mathbf{c_i})$ that indicate the linguistic property, topic of discussion, metadata about the author, and the community in which the comment occurred, respectively. Note that each $\mathbf{s_i}, \mathbf{u_i}, \mathbf{c_i}$ can be a vector including a series of subfactors. 

A common approach for estimating the change in a property is to simply calculate the mean of the $\hat{l}_i$'s over time~\cite{chandrasekaran2020topics,jang2022covid, zhou2023covid}. This approach is typically accompanied by an interpretation that implicitly assumes the $\mathbf{s_i}, \mathbf{u_i}, \mathbf{c_i}$ remain static across users over time. In other words, one assumes that the overall change was driven by user change rather than a shift in the composition of who is on the platform or where the discourse occurs. However, we argue that this is overly simplistic. Just as identical summary statistics can be generated by qualitatively different underlying data distributions, as in Anscombe's quartet, widely different mechanisms can drive the same measured changes in online discourse. 

We consider three main mechanisms that can drive overall changes in online stance towards a topic. 

\begin{enumerate}
    \item \xhdr{Content drift} Each post on a platform is defined by its content. A macro-shift in stance can be explained either by changes in distribution for the topics of discussion, and/or a shift in the stances on the individual topics.
    \item \xhdr{Generational drift} The work of Waller and Anderson~\shortcite{waller2021quantifying} found that polarization on Reddit was not a result of polarization of individual users, but instead a product of fringe new cohort of radical users. Since users are often anonymous on these platforms, we distinguish user features by their cohort. 
    \item \xhdr{Behavioral drift} Context forms both online and offline discussion. How users speak in one group may differ widely from another group. Overtly this can be the result of community banning or inception. More subtle shifts can take place with specific issues being politicized across one side. 
\end{enumerate}

\subsection{Counterfactual Modelling}
To measure the impact of these three factors on driving changes in opinion, we design a simple counterfactual modelling setup. 
While more traditional methods like linear regression and t-tests quantify the predictive power of certain independent variables over the other or compare the difference between two groups of data, it does not make a causal claim. In comparison, counterfactual scenarios allows a more faithful approach the question ``what drove overall stance change?'' by simulating what would have happened if only certain content, generational, and behavioural drifts had happened. 
Our setup relies on the interplay of proportion and stance and enables us to measure the impacts of the different meta-factors' proportions and stances in isolation from other influences.

Let $S$, $U$, and $C$ denote all the possible values that $\mathbf{s}$, $\mathbf{u}$, and $\mathbf{c}$ can take on, respectively, across the entire platform $P$. Let $M \subset S \times U \times C$ denote a subset of the metadata. Let $\text{P}_t(M)$ be the proportion of comments in a specific quarter $t$ that include the meta-factors in $M$ and $\text{L}_t(M)$ be the average stance for that group in quarter $t$. The total stance on the platform for a given time $t$ is equal to 

\begin{align*}
    L(t) &= \sum_{(\mathbf{s},\mathbf{u},\mathbf{c})}{\text{P}}_t(\{(\mathbf{s},\mathbf{u},\mathbf{c})\}) \cdot {\text{L}}_t(\{(\mathbf{s},\mathbf{u},\mathbf{c})\})
\end{align*}

We then denote $\bar{\text{P}}(M)$ as the average proportion for comments with conditions in $M$ over all time, and similarly define $\bar{\text{L}}(M)$.
For any set $M$, we can consider four different counterfactual scenarios. First, we may be interested the effect of both the proportion and the stance on $L(t)$. Second, we may be interested in the effect of only the proportion on $L(t)$. Third, we may be interested in the effect of only the stance on $L(t)$. Fourth, we may be interested in keeping both the proportion and the stance fixed across time, to control for the effect of the elements in $M$ on $L(t)$.
Let $\mathcal{M}_1, \mathcal{M}_2, \mathcal{M}_3, \mathcal{M}_4$ be sets where each element in the set is a subset of $S \times U \times C$. $\mathcal{M}_1,$ represents the metadata for which we are interested in both the proportion and the stance, $\mathcal{M}_2$ represents the metadata for which we are interested in only the proportion, $\mathcal{M}_3$ represents the metadata for which we are interested in only the stance, and $\mathcal{M}_4$ represents the metadata for which we are interested in neither the proportion nor the stance.
Note that $(\bigsqcup \mathcal{M}_1) \sqcup (\bigsqcup \mathcal{M}_2) \sqcup (\bigsqcup \mathcal{M}_3) \sqcup (\bigsqcup \mathcal{M}_4) = S \times U \times C$, which represents the entire social media platform $P$.

Now we consider the counterfactual world as 
\begin{align*}
    L(t) &= \sum_{M_1 \in \mathcal{M}_1}{\text{P}}_t(M_1) \cdot {\text{L}}_t(M_1) \\
    &+ \sum_{M_2 \in \mathcal{M}_2} {\text{P}}_t(M_2) \cdot \bar{\text{L}}(M_2) \\
    &+ \sum_{M_3 \in \mathcal{M}_3} \bar{\text{P}}(M_3) \cdot {\text{L}}_t(M_3) \\
    &+ \sum_{M_4 \in \mathcal{M}_4} \bar{\text{P}}(M_4) \cdot \bar{\text{L}}(M_4)
\end{align*}

To clarify the formalization of our counterfactual modelling setup, we illustrate two examples.  First, suppose we are interested in the how well the proportion of topics of discussion can predict the stance of the platform for a given time. In this setting, we have $\mathcal{M}_2 = \{\{s\} \times U \times C \ | \ s \in S\}$, since we are interested in varying the proportion of the subtopics. Also, $\mathcal{M}_1 = \mathcal{M}_3 = \mathcal{M}_4 = \emptyset$, since we are not interested in the effect of the stances or the proportions of any other meta-factors on overall stance. 
Second, suppose that we are interested in how well the stance of the users can predict the stance of the platform for a given time. In this setting, we have $\mathcal{M}_3 = \{S \times \{u\} \times C \ | \ u \in U\}$, since we are interested in varying the stance of the users. Also, $\mathcal{M}_1 = \mathcal{M}_2 = \mathcal{M}_4 = \emptyset$, since we are not interested in the effect of the stances or the proportions of any other meta-factors on overall stance. 

\section{Empirical Methods}
\label{sec:empirical}

Guided by this formalization, we develop an empirical method capable of disambiguating between mechanisms of online opinion change. In this paper, we apply our method to studying public opinion change about UBI on Reddit.

\subsection{Data}

Our dataset consists of all Reddit comments up to June 2022 containing the phrase ``basic income'' (case-insensitive) or ``UBI'' (case-sensitive). The first phrase is used in existing literature \cite{gielens2022more}, and the second phrase is added due to its observed frequency of use. 
More permissive approaches for dataset creation (e.g., snowball sampling) led to greatly reduced precision; see the Appendix for details. 

We filtered the initial dataset of 1.39M comments down to 1.22M comments across 9,827 unique Reddit communities (\emph{subreddits}) and 340K unique authors (filtering strategy described in the Appendix \ref{appendix:data}). 
We find that many UBI comments occur in subreddits that frequently discuss political and economic topics. The top 10 most represented subreddits in our dataset are r/Futurology, r/politics, r/BasicIncome, r/YangForPresidentHQ, r/AskReddit, r/worldnews, r/canada, r/antiwork, r/news, and r/Economics. Comments in these 10 subreddits account for 42.4\% of all comments in our dataset.

To validate the quality of our dataset, we selected a random sample of 500 comments and found that $<$ 1\% of comments were unrelated to UBI. Since the number of UBI comments in a subreddit follows a long-tailed distribution, we also tried a stratified sampling approach to validate our dataset. We divided the subreddits into 10 groups by first ordering them by their number of UBI comments, and then taking the subreddits from the 0-10th percentile, 10-20th percentile, and so forth.
Then, we randomly sampled 50 comments from each group and found that only $<$ 6\% of comments were unrelated to UBI.

\subsection{Methods}
We design three methods to augment the set of comments with further context on \emph{what} is being discussed (content analysis), \emph{who} is discussing it (user analysis), and \emph{where} the discussions are taking place (community analysis). 

\xhdr{Stance Detection} We labelled each comment in our dataset with the commenter's attitude towards UBI (supportive, neutral, or against). Table~\ref{tab:stance_examples} contains three comments labelled with their stances.

\begin{table}[t!]
    \small
    \centering
    \begin{tabular}{c|p{5cm}}
    \toprule
         Stance & Example Comment \\
         \midrule
         Supportive & ``This why Universal Basic Income is not a bad idea. We need to figure it out because in 30 years there won't be any jobs like jobs that exist today.'' (in r/technology) \\
         Neutral & ``Why do you say UBI is immoral?'' (in r/CapitalismVSocialism) \\
         Against & ``The scaled up UBI program would cost 16 billion, roughly the same sum that they are using right now on social security and welfare programs. The idea was that UBI would replace the welfare programs, which is also part of why I (and most communists) are opposed to it'' (in r/PoliticalCompassMemes)\\
    \bottomrule
    \end{tabular}
    \caption{Examples of comments labelled supportive, neutral, and against.}
    \label{tab:stance_examples}
\end{table}
Labelling was done through a two-step process. First, we prompted GPT-4 to label a comment with the commenter's attitude towards UBI. The prompting strategy is described further in the Appendix\ref{appendix:stance}. After labelling 10,000 comments in this way, we trained a smaller-scale RoBERTa-large model on these labels.

To assess the accuracy of the two steps, one of the co-authors manually labelled a random sample of 100 comments with their stance. 
Twenty of these labels were then further validated by two other co-authors.
We discarded 5 comments that were not about UBI. 
Comparing the GPT-4 labels with the manual labels, we get an accuracy of 0.768 and a macro-F1 of 0.766. Comparing the model labels with the manual labels, we get an accuracy of 0.611 and a macro-F1 of 0.594.
Note that for some comments, more than one stance was reasonable (usually supportive / neutral, or neutral / against) (see the Appendix\ref{appendix:stance} for an example for when this was the case).
Notably, for both the GPT-4 and model labels, only 2 and 6 of the 95 comments in the random sample, respectively, were comments that were misclassified as supportive when the comment was against or vice versa.

\xhdr{Topic Creation}
We first apply latent Dirichlet allocation (LDA) topic modelling~\cite{blei2003latent} to our dataset to computationally extract themes of discussion. 
The details of how we preprocessed the data and ran LDA are described in the Appendix. 
Using the topic model with 19 topics, we name each of the topics by examining the top 10 words and top 10 comments.
We manually validate comments in these topics to ensure that the topics are coherent. We decide to discard 4 noisy topics, leaving us with a total of 15 (see Table~\ref{tab:topics_descriptions}).
\begin{table*}[t!]
    \scriptsize
    \centering
    \begin{tabular}{c|p{12.5cm}}
    \toprule
        Subtopic Name & Description \\
        \midrule
        Living costs & Comments in this class could discuss UBI's impact on the housing market, rent, or the prices of essential goods. They could also discuss the supply and price of land, UBI's impact on landowners and renters, or whether UBI incentivizes people to move. \\
        Data analysis and research & Comments in this class could discuss UBI experiments, studies, and trials, or contain links to other UBI-related resources, such as news articles. They could also debate whether there is evidence on whether UBI works or not. \\ 
        Education and family & Comments in this class could discuss UBI's impact on different genders, parents, or child care. They could also discuss family planning. \\
        Non-UBI government welfare programs &  Comments in this class could discuss government welfare programs that are commonly seen as alternatives to UBI, such as social security, food stamps, or unemployment insurance. They could compare UBI with these alternatives or discuss whether UBI will replace these alternatives.\\
        Budget and cost & Comments in this class could discuss the monetary cost and government spending required to implement UBI, and whether is it affordable or not. \\
        Economic systems &  Comments in this class could discuss economic systems such as socialism, communism, and capitalism. They could also debate whether UBI is socialist or not. \\
        Labor wages and work conditions & Comments in this class could discuss UBI's impact on wages and the incentive to work. They could also discuss the employee's bargaining power in the employee - employer relationship. They could also discuss the kinds of jobs people are free to choose from. \\
        Public services and healthcare & Comments in this class could discuss other public policies that are relevant today, such as healthcare, drug policy, prison reform, public education, and student debt. \\
        Money and inflation & Comments in this class could discuss banks or the Federal Reserve. They could also discuss the supply of money, whether printing money will cause inflation, or whether UBI will cause inflation. \\
        Politics and elections &  Comments in this class could discuss different candidates and parties in a political election, the results of an election, or voting. \\
        Global affairs &  Comments in this class could discuss non-US countries or compare first and third-world countries. They could also discuss immigration, foreign relations, or war. \\
        Automation and jobs &  Comments in this class could discuss job loss through automation and technological progress, and whether UBI is a solution to this problem. \\
        Taxes &  Comments in this class could discuss different types of taxes, such as progressive and regressive taxes, Negative Income Tax (NIT), or Value Added Tax (VAT). \\
        Political affiliations & Comments in this class could compare views on specific political issues between political left and right. They could also discuss people and parties that belong to the political left vs right. They could also contain the commenter's political views.\\
        Businesses and profit & Comments in this class could discuss UBI's impact on companies, their stocks, and their profits. They could also discuss UBI's impact on small vs large businesses. \\
        \bottomrule
    \end{tabular}
    \caption{Subtopic name and a description of comments belonging to the subtopic.}
    \label{tab:topics_descriptions}
\end{table*}

\xhdr{Topic Extraction}
Although LDA is a useful method to gain a rough picture of all the topics in a dataset, its bag-of-words nature risks losing the semantic relationships between the individual words in a comment. 
To account for this, we again use a two-step method to train an encoder-only classifier. First, we designed a prompting strategy for \turbo{} to extract potentially multi-label topics from a comment. The prompt consisted of five choices: the four most probable topics determined by LDA, and ``None of the above.'', and is closely described in the Appendix\ref{appendix:topics}.
After labelling 10,000 comments in this way, we finetuned a \texttt{e5-base} model~\cite{wang2022text} on these labels. 

To assess the accuracy of the two steps, one of the co-authors manually labelled a random sample of 100 comments with the LDA topics they pertain to. We labelled a comment with ``None of the above'' if the comment did not pertain to any subtopic or was not about UBI. 
Twenty of these labels were then further validated by one other co-author.
Comparing the \turbo{} labels with the manual labels, we get a macro-F1 of 0.412 and a micro-F1 of 0.455. Comparing the model labels with the manual labels, we get a macro-F1 of 0.398 and a micro-F1 of 0.450. 
While the performance of our model could affect the quality of our results, we believe that our model reasonably captures the subtopics contained within a comment, since we had a total of 16 possible labels (the 15 labels provided by LDA and ``None of the above''), and each comment can contain 1-4 of these labels. Furthermore, our model outperforms in both macro-F1 and micro-F1 compared to the maximum F1 scores we can get by using only LDA (which is a macro-F1 of 0.393 and micro-F1 of 0.417). The details of this calculation are in the Appendix.

\xhdr{User Analysis} We classify all Reddit users by the year that they made their first comment on the platform and call each grouping of users a ``cohort''. We then study how the changing distributions of cohort contributions over time affects the overall stance towards UBI on Reddit. 

\xhdr{Community Analysis} To gain a behavioral understanding of where the UBI discourse on Reddit occurs, we use community embeddings, an established technique to study online social media through a behavioral lens~\cite{waller_anderson_2019}.
The embeddings permit simple vector algebra to create meaningful semantic dimensions like partisan-leaning, age, sociality, and affluence~\cite{waller2021quantifying}. We project the Reddit communities onto the four social dimensions to get social-context for each of the communities (see the Appendix for a detailed description of the community embeddings).

We then categorize each community into five bins along each social dimension. For example, for the partisan dimension, we order all Reddit communities in the embeddings by their partisan score, then assign each community a partisan percentile score (0\% means that the community is the most left-wing community; 100\% means that the community is the most right-wing community. We then divide the communities into five equally-sized bins, defining left-wing, center-left, center, center-right, and right-wing communities as communities with percentile score from 0-20\%, 20-40\%, 40-60\%, 60-80\%, and 80-100\%, respectively. 

\xhdr{Comparing Time Series} To perform the content, user, and community analyses using our counterfactual model, we need to compare the time series of stance generated by the counterfactual model to the actual changes in stance that we observe. There are many ways of quantifying the similarity between two time series, including Pearson correlation, Euclidean distance, and Dynamic Time Warping (DTW)~\cite{kianimajd2017comparison}.
Each measure is sensitive to different types of time series offsets; for instance, the Euclidean distance is sensitive to value offsets.
For the main results of our work, we choose to compare the time series using Pearson correlation. We include results from two other measures---Euclidean distance and DTW---in the Appendix.

\section{Results}
We now apply the conceptual model from before on UBI-related Reddit comments. 
We consider the content-, user-, and community-level characteristics of UBI comments---the potential drivers of overall stance change---in isolation. 
For each potential driver, we apply counterfactual thinking to model what would have happened if only shifts in stance had occurred (which we refer to as the first counterfactual scenario) v.s. if only shifts in distribution had occurred (which we refer to as the second counterfactual scenario).
For example, on the content-level, stance change be either driven by a shift in stances across various subtopics of UBI or a shift in distribution around which subtopics are being discussed.
We compare the counterfactual scenarios with the empirical stance, which we define as the changes we observe in Figure~\ref{fig:overall-stance}.
We perform this comparison by taking the Pearson $r$ correlation coefficient between the stances over time produced by the counterfactual and the empirical stance (we explore measures other than Pearson correlation in the Appendix).
If, for a driver, the Pearson correlation between the first counterfactual and the empirical stance is higher than the Pearson correlation between the second counterfactual and the empirical stance, then for this driver, stance changes are more responsible for overall stance change than distribution shifts.
Overall, our methodology allows us to explore how content-, user-, and community-level factors contributed to the empirical stance. 

\xhdr{Content-level}
As mentioned above, an observed drop in stance may either be due to an actual shift in stance (corresponding to the first counterfactual scenario) or a shift in what people discuss, moving to more negative topics (corresponding to the second counterfactual scenario). 

\begin{figure}[t]
    \centering
    \includegraphics[width=7.7cm]{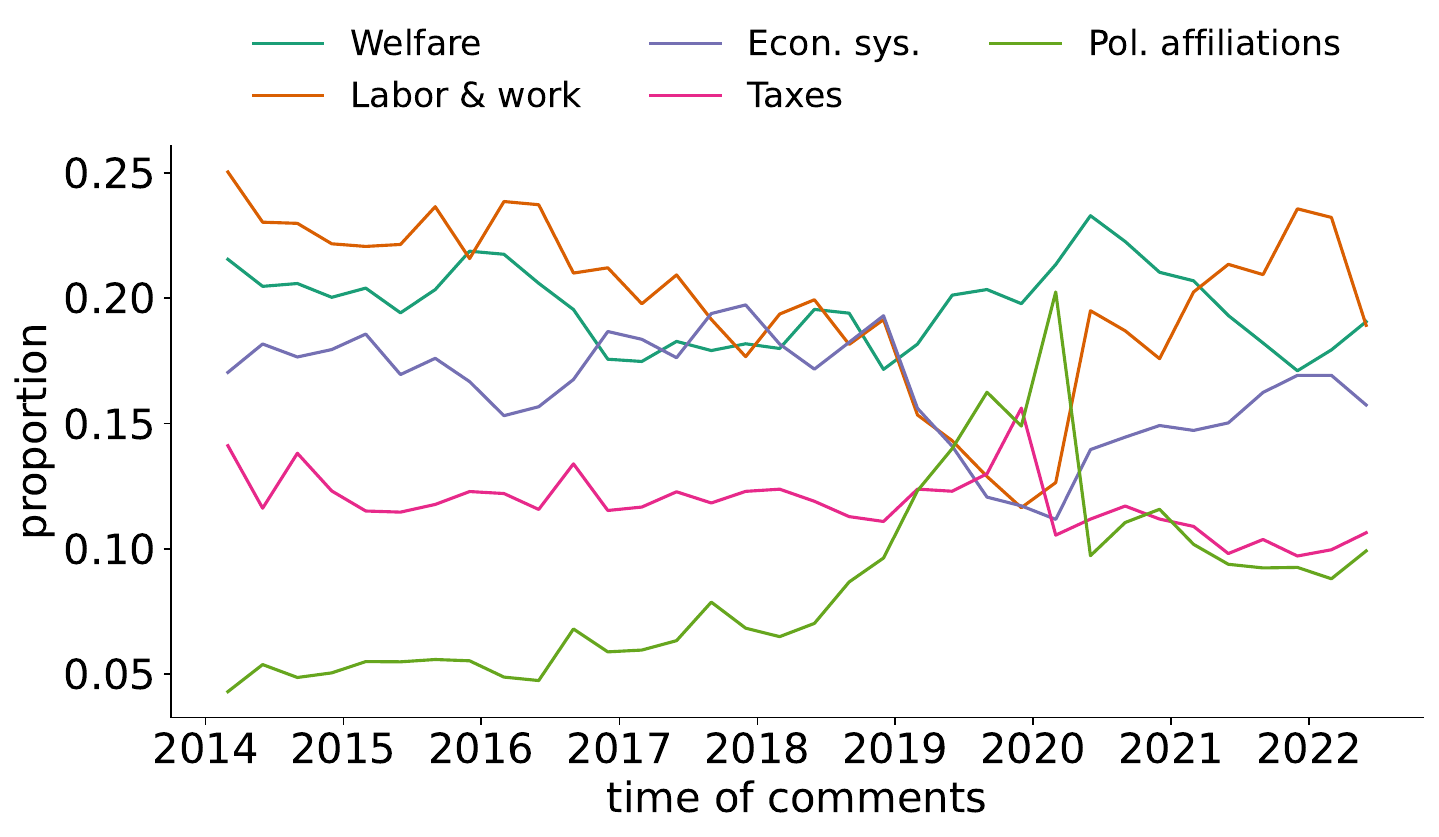}
    \caption{Proportion of comments that pertain to the largest five subtopics. 67.4\% of all the UBI comments pertain to one of these five subtopics.}
    \label{fig:subtopic-proportion}
\end{figure}

\begin{figure}[t]
    \centering
    \includegraphics[width=7.7cm]{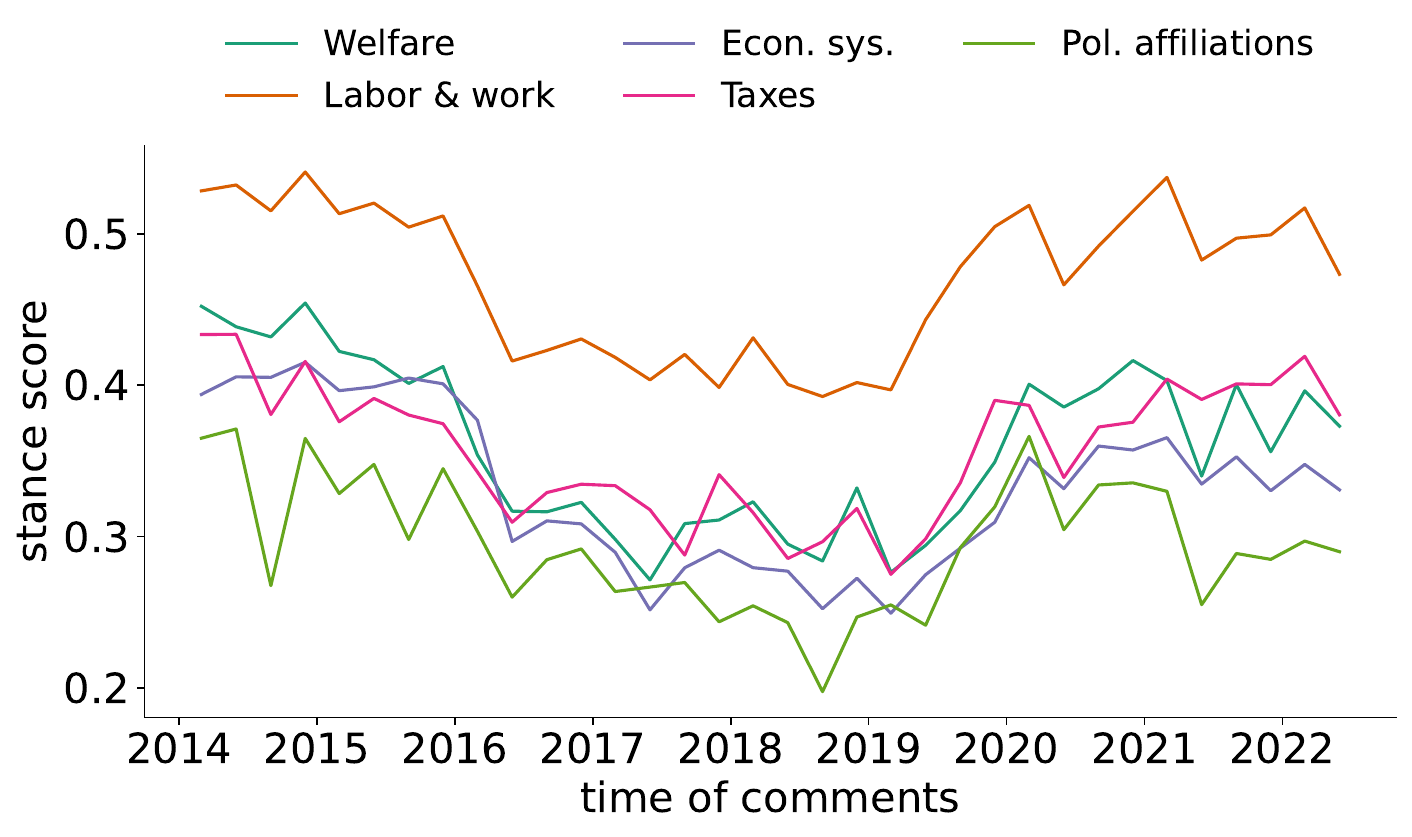}
    \caption{Average stance of UBI comments in the largest five subtopics.}
    \label{fig:subtopic-stance}
\end{figure}

\begin{figure*}
    \centering
    \includegraphics[width=14.8cm]{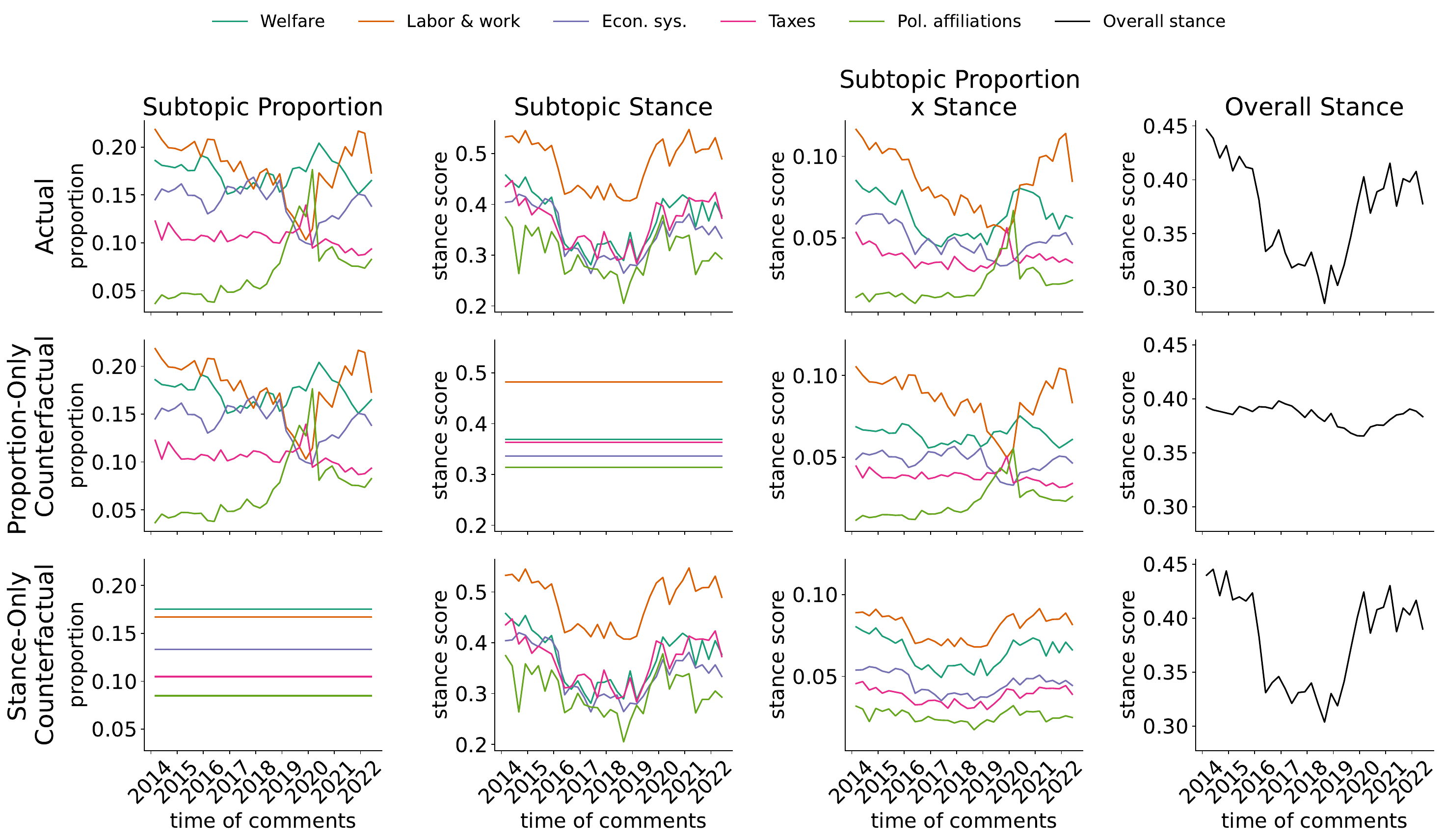}
    \caption{Actual, proportion-only counterfactual, and stance-only counterfactual scenarios. The top row illustrates how the subtopic proportions and stances and the overall stance changed over time. The middle row illustrates the counterfactual scenario where we keep the stances constant, and let the proportions vary. The bottom row illustrates the counterfactual scenario where we keep the proportions constant, and let the stances vary.}
    \label{fig:subtopic-3-by-4}
\end{figure*}
We observe how distributions across and stances on UBI subtopics have changed from 2014 to 2022.  
The results are shown in Figure~\ref{fig:subtopic-proportion} and Figure~\ref{fig:subtopic-stance}. We see that subtopic stance appears to be more correlated with overall stance shifts in UBI than subtopic proportions. 
The two counterfactual scenarios for the top 5 subtopics are shown in Figure~\ref{fig:subtopic-3-by-4}.
To quantify this observation, we observe that the Pearson $r$ correlation coefficient between the first counterfactual and the empirical stance is only $0.126$ ($p = 0.476$) while the Pearson $r$ correlation coefficient between the second counterfactual and the empirical stance is $0.982$ ($p = 9.281 \cdot 10^{-25}$).
These calculations demonstrate that subtopic stance is more responsible for stance shifts in UBI than proportion of subtopics.
In other words, stances are changing within most subtopics, rather than there being primarily negative and positive subtopics. 
In the Appendix, we show that with alternative measures, we get the same overall result regarding the subtopics.

\begin{figure}[t]
    \centering
    \includegraphics[width=7.7cm]{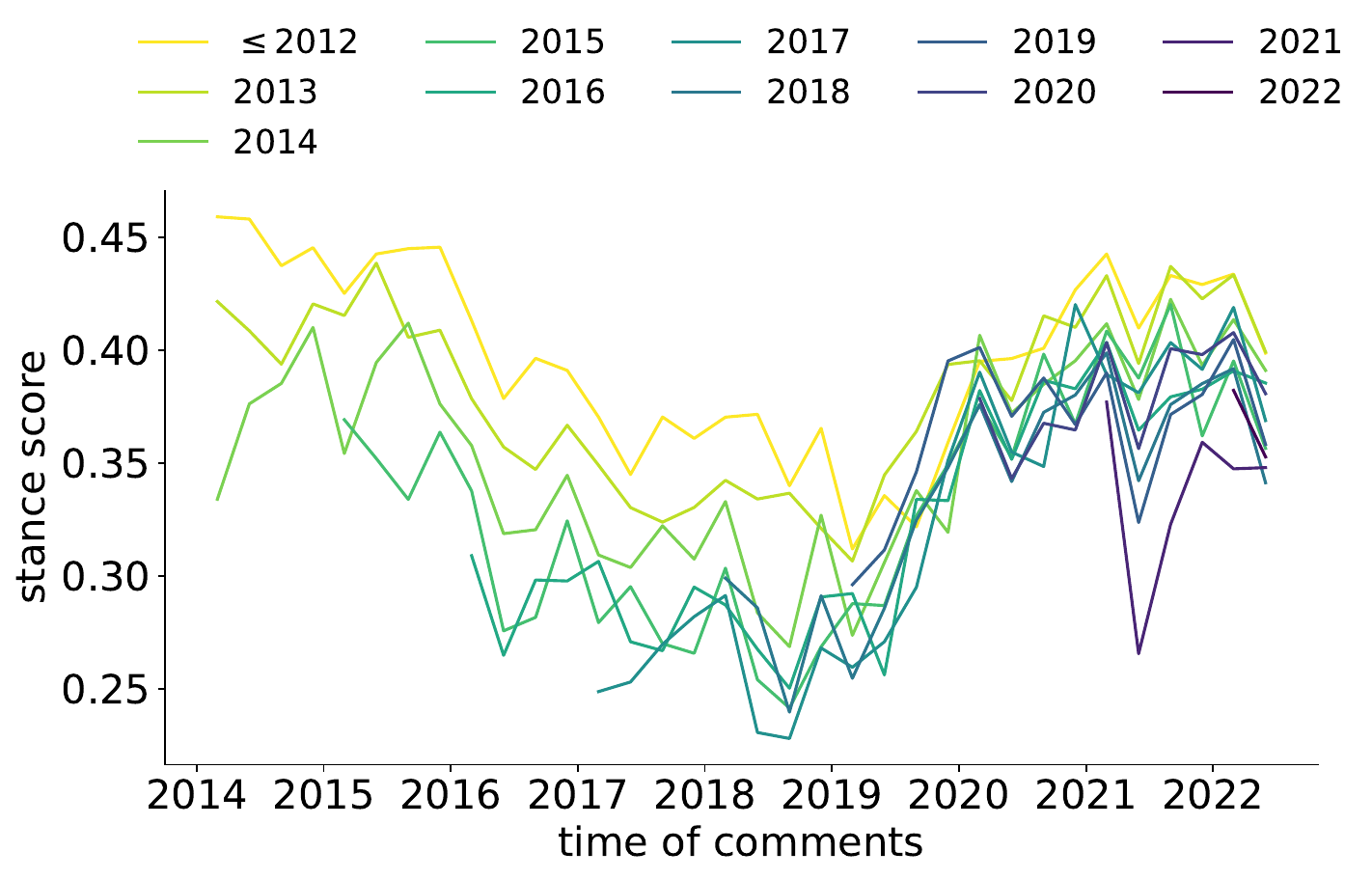}
    \caption{Average stance on UBI comments in cohorts from 2012-2022.}
    \label{fig:cohort-stance}
\end{figure}

\xhdr{User-level analysis}
In Figure~\ref{fig:cohort-stance}, we observe how stances on UBI change over time for different cohorts. 

\begin{figure}[t!]
    \centering
    \includegraphics[width=7.7cm]{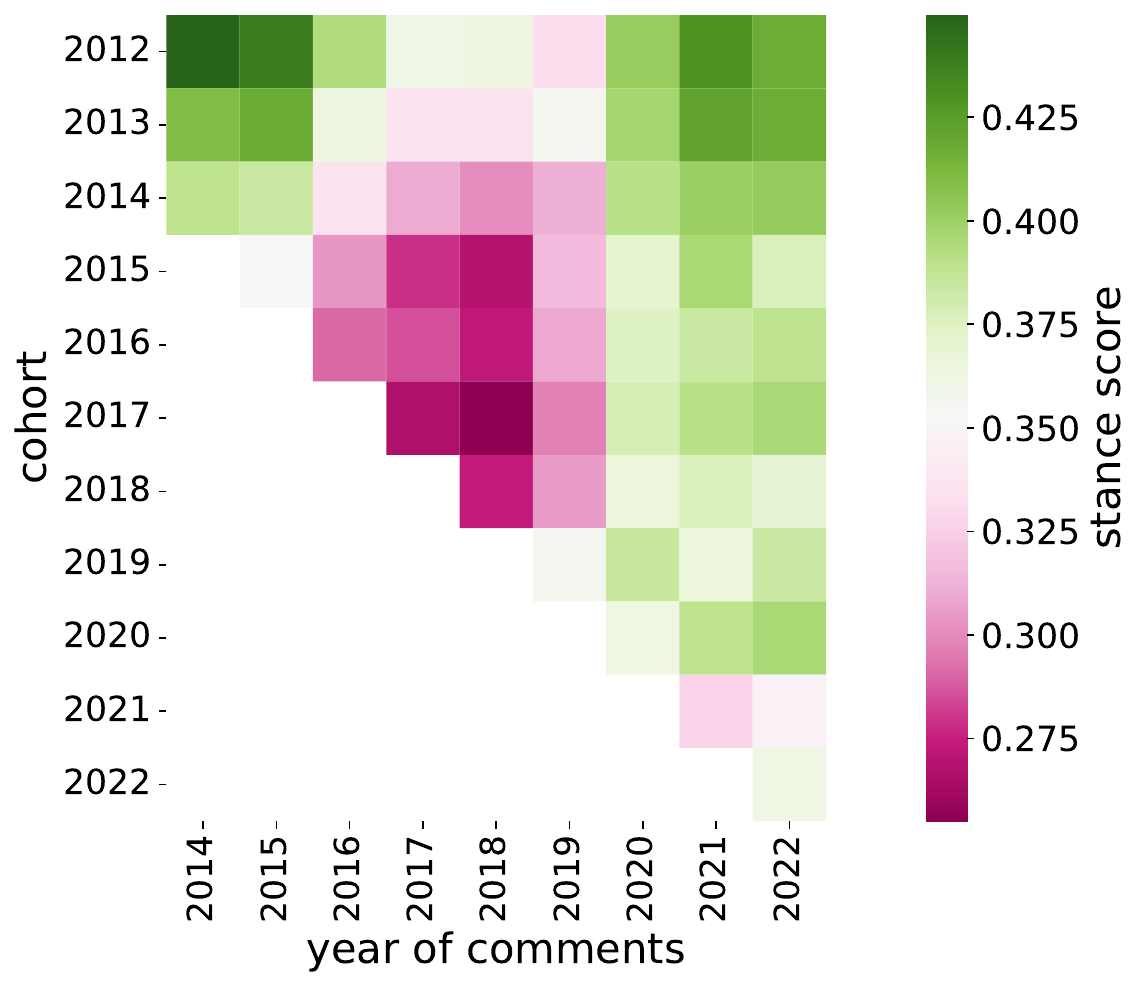}
    \caption{Average stance on UBI comments in cohorts from 2012-2022. The y-axis represents the cohort that made the comments. The x-axis represents the year the comments was made. The colour in each cell represents the average stance of a given cohort's UBI comments in a given year.}
    \label{fig:cohort-heatmap}
\end{figure}

We observe two effects. First, the individual cohorts change their stances over time, and in particular become negative from June 2016 to March 2019. Second, newer cohorts are more negative about UBI than older cohorts. 
Figure~\ref{fig:cohort-heatmap} further highlights these observations.

We again perform the counterfactual analyses on distribution vs. stance. An illustration of this calculation is in Appendix Figure~\ref{fig:appendix-cohort-3-by-4}. 
Note that the implementation of the counterfactual analysis is different for the user-level analysis compared with the content-level and community-level analyses because not every cohort exists in every year. Thus, as illustrated in the bottom row of Appendix Figure~\ref{fig:appendix-cohort-3-by-4}, we renormalize the proportions for each cohort in each quarter so that the proportions add to 1.
The Pearson $r$ correlation coefficient between the first counterfactual and the empirical stance is $0.791$ ($p = 2.603 \cdot 10^{-8}$), while for the second it is $0.994$ ($p = 5.774 \cdot 10^{-32}$).
Thus, stance shift within each cohort is more responsible for stance shifts in UBI than proportion of comments made by each cohort. In the Appendix, we show that with alternative measures, we get the same overall result regarding the cohorts.

\begin{figure}[t!]
    \centering
    \includegraphics[width=7.7cm]{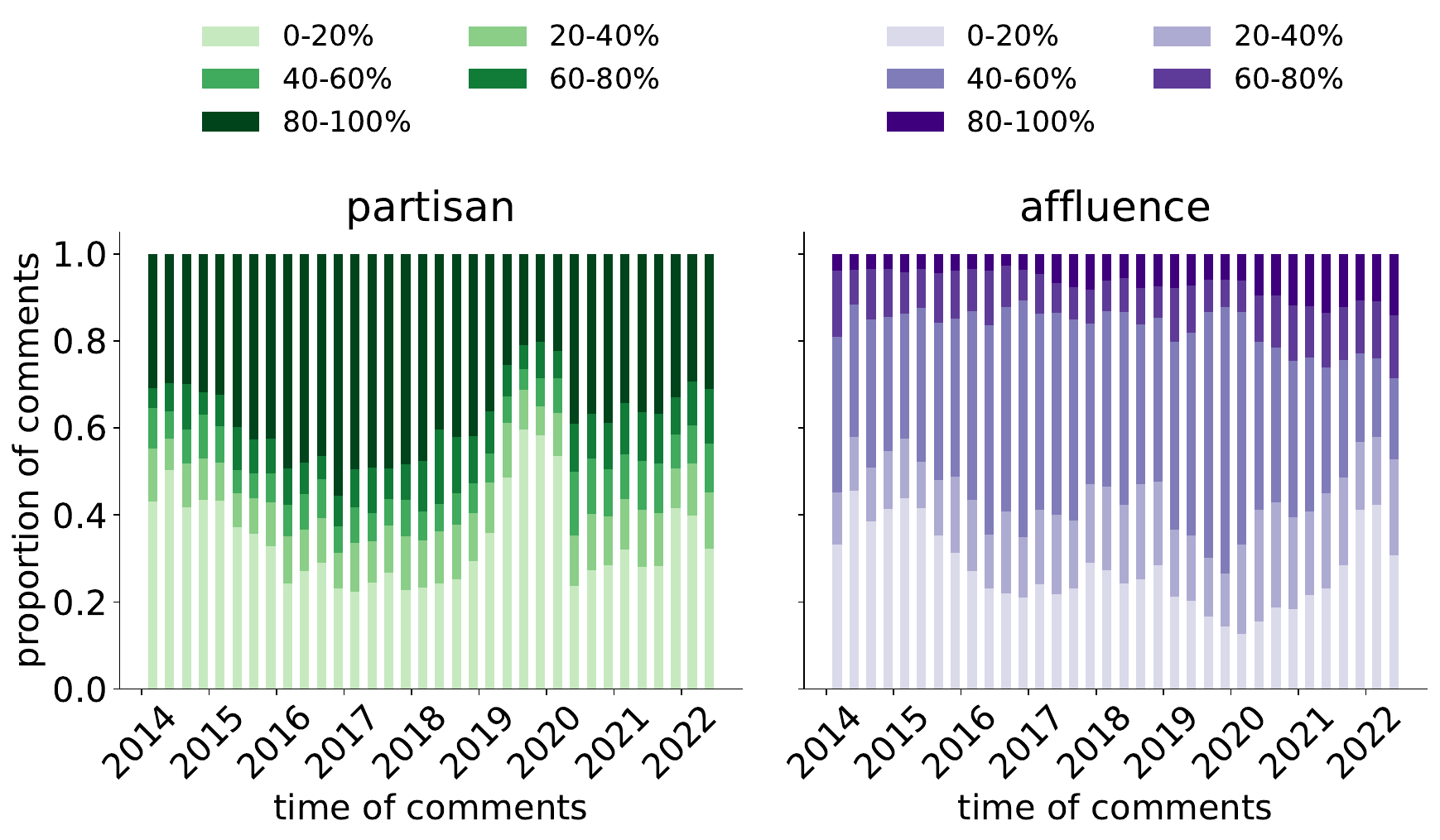}
    \caption{Proportion of UBI comments made by communities along the partisan and affluence dimensions. Lighter colours represent left-wing or less affluent communities, and darker colours represent right-wing or more affluent communities.}
    \label{fig:dimension-proportion}
\end{figure}

\begin{figure}[t!]
    \centering
    \includegraphics[width=7.7cm]{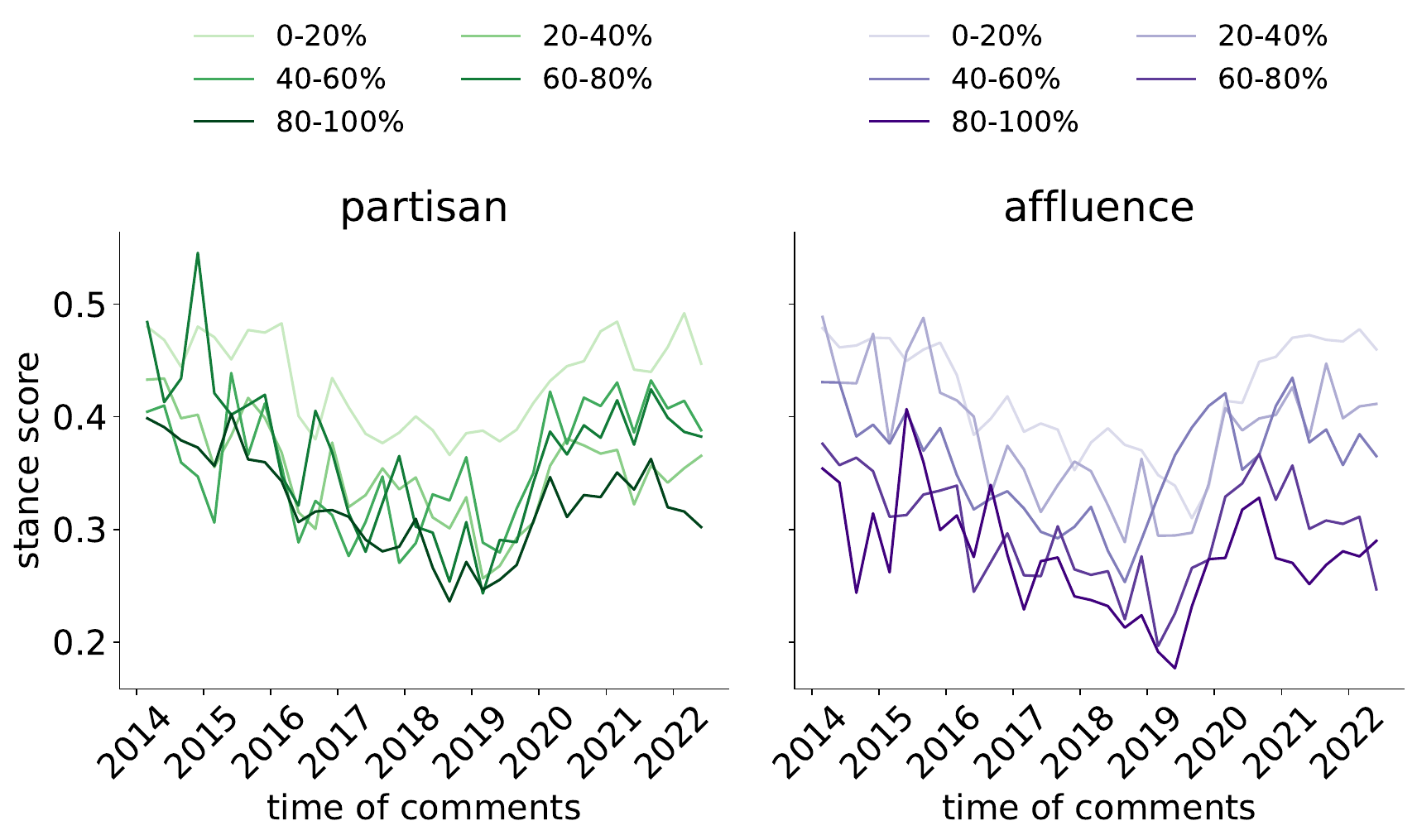}
    \caption{Average stance on UBI comments along the partisan and affluence dimensions. Lighter colours represent left-wing or less affluent communities, and darker colours represent right-wing or more affluent communities.}
    \label{fig:dimension-stance}
\end{figure}

\xhdr{Community-level analysis}
Using the bucketed scores for each Reddit community, we observe how the distribution of UBI comments changes over time. The results for the partisan and affluence dimensions are illustrated in Figure~\ref{fig:dimension-proportion}. We also observe how stances on UBI change over time in the different bins; the results for the partisan and affluence dimensions are illustrated in Figure~\ref{fig:dimension-stance}. 
We also include the corresponding results for the age and gender dimensions in Figure~\ref{fig:appendix-age-gender-proportion} and Figure~\ref{fig:appendix-age-gender-stance}. An overwhelming majority of UBI comments occurred in older and masculine communities, leading to noise in the smaller bins in for the age and gender dimensions.


\begin{figure*}
    \centering
    \includegraphics[width=14.8cm]{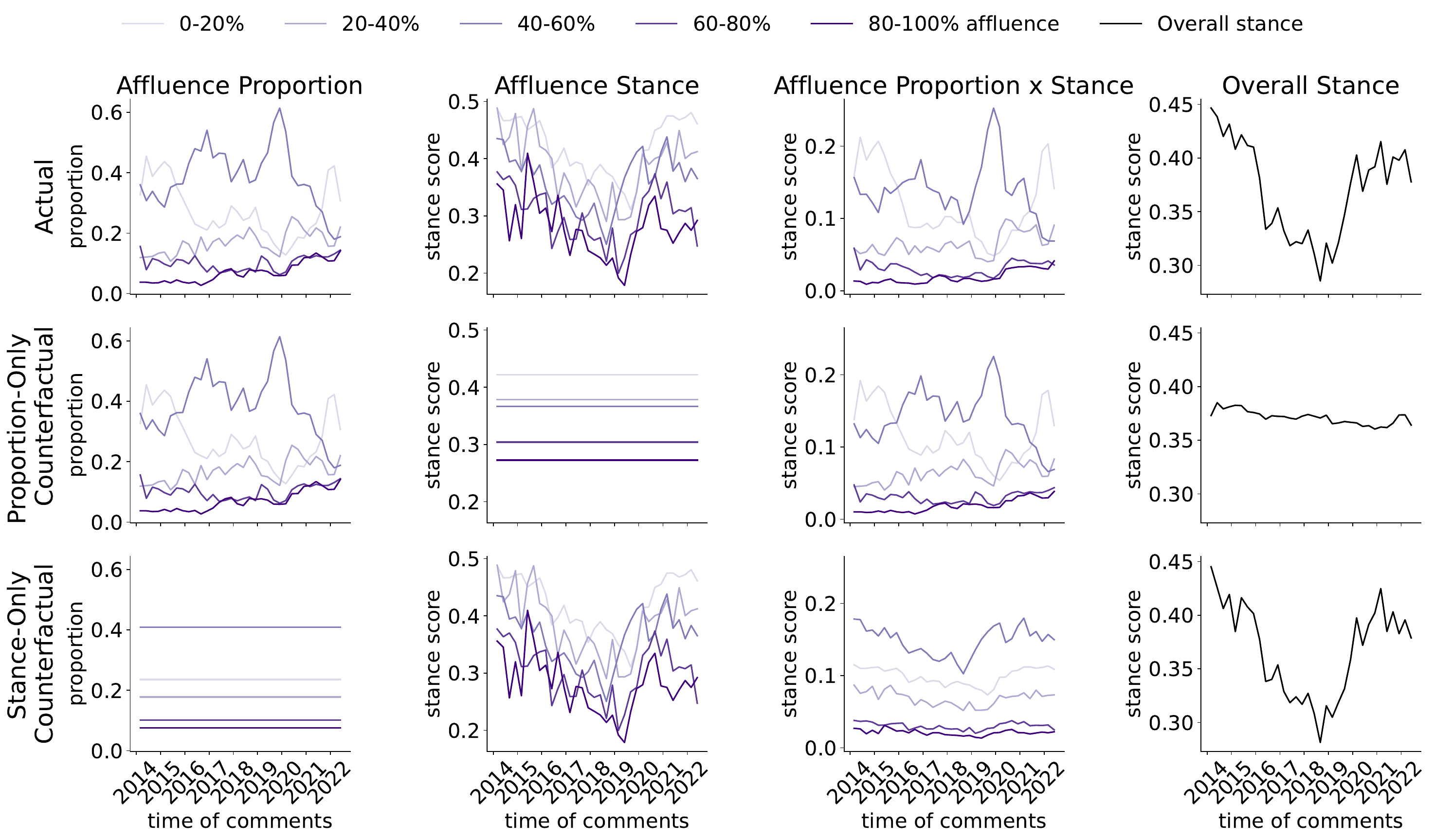}
    \caption{Actual, proportion-only counterfactual, and stance-only counterfactual scenarios. The top row illustrates how the proportions and stances in each affluence bin and the overall stance changed over time. The middle row illustrates the counterfactual scenario where we keep the stances constant, and let the proportions vary. The bottom row illustrates the counterfactual scenario where we keep the proportions constant, and let the stances vary.}
    \label{fig:affluence-3-by-4}
\end{figure*}

We again perform the counterfactual analyses on distribution v.s. stance. An illustration of this calculation is in Appendix Figure~\ref{fig:appendix-partisan-3-by-4} and Figure~\ref{fig:affluence-3-by-4}.

For the partisan dimension, the Pearson $r$ correlation coefficient between the first counterfactual and the empirical stance is only $0.462$ ($p = 0.006$), while for the second it is $0.971$ ($p = 1.643 \cdot 10^{-21}$).
For the affluence dimension, the Pearson $r$ correlation coefficient between the first counterfactual and the empirical stance is only $0.320$ ($p = 0.065$) while the Pearson $r$ correlation coefficient between the second counterfactual and the empirical stance is $0.984$ ($p = 1.945 \cdot 10^{-25}$).
These calculations demonstrate that for both the partisan and affluence dimensions, stance shift within each bin is more responsible for stance shifts in UBI than proportion of comments made in each bin.
In other words, in determining overall stance on UBI, it matters less how much UBI discussion occurred left- v.s. right-wing and less v.s. more affluent; what matters more is the stance changes that occurred \emph{within} left- v.s. right-wing and less v.s. more affluent communities. In the Appendix, we show that with alternative measures, we get the same overall result regarding the partisan and affluence dimensions.

Finally, we observe an polarization in the affluence dimension starting in 2019. Namely, in the plot showing the actual stance Figure~\ref{fig:dimension-stance} the difference between the most affluent communities (in dark purple) and the least affluent communities (in light purple) grows following 2019. 
Interestingly, such a polarization event does not occur for the partisan dimension.
This has important implications for people interested in UBI as it demonstrates how opinions surrounding UBI are diverging across some groups, but not others.

\xhdr{Combining the analyses}
In this section, we combine our analyses on the subtopics, cohorts, and communities to determine which of the three best explains overall stance change on UBI. 
We define the loss as sum of the absolute values of the difference between the empirical stance and the predicted stance under a counterfactual world from January 2014 to June 2022.
We calculate this loss for each of the counterfactual worlds we explored in the previous sections. The results are illustrated in Table~\ref{tab:all-counterfactuals}.
\begin{table}
    \centering
    \begin{tabular}{l|c|c}
    \toprule
         Varying Property & Pearson $r$ & Loss \\
         \midrule
         Subtopic Proportion & 0.126 & 1.315 \\
         Subtopic Stance & 0.982 & 0.361 \\
         Cohort Proportion & 0.791 & 1.640 \\
         Cohort Stance & 0.994 & 0.172 \\
         Partisan Proportion & 0.462 & 1.219 \\
         Partisan Stance & 0.971 & 0.300 \\
         Affluence Proportion & 0.320 & 1.292\\
         Affluence Stance & 0.984 & 0.227 \\
    \bottomrule
    \end{tabular}
    \caption{The Pearson $r$ and losses between each counterfactual world and the empirical stance. The loss is the sum of the absolute values of the difference between the empirical stance score and the predicted stance score from January 2014 to June 2022.}
    \label{tab:all-counterfactuals}
\end{table}

Overall, we see that stance changes within subtopics, cohorts, and dimensions are better at describing overall stance change than proportion changes.
This means that UBI is discussed both negatively and positively in across most subtopics, in both left- and right-wing and less and more affluent communities, and across different cohorts.
This suggests that stances on UBI are constantly changing on the micro-level. Our analyses show that different topics, users, and communities can quickly become more positive or negative about UBI.
The counterfactual world that best explains the overall stance change is the one in which we fix the proportions across the different cohorts, and let the stances vary.
\section{Discussion}
Public discourse is a key driver of societal change, and it is increasingly taking place online. Understanding the common debate about public policies helps us make better collective decisions. While traditional methods of measuring public opinion can capture macro-level trends, they have limited ability to provide fine-grained insight into whether changes in public opinion are occurring along important social lines and especially what mechanisms are driving opinion change. As changes in public sentiment due to population drift versus those due to genuine changes in opinion call for different policy responses and interventions, we need higher-resolution and finer-grained methods to measure the mechanisms of public opinion change.

In this work, we contribute a method of measuring online public opinion change on a particular topic in a high-resolution and fine-grained way. Our technique can distinguish  between qualitatively different mechanisms that can drive the same overall changes in public opinion, and which all might call for different policy responses.  
A surge in negative sentiments on a platform, for instance, might stem from various factors--such as an influx of pessimistic users, rising and fading popularity of positive or negative communities, or a transformation in the discussion framing itself.
Without a granular understanding of the forces that drive these changes, our ability to derive actionable insights is limited. To bridge this gap, our work presents a conceptual framework for studying subtle shifts that drive broader opinion change. 
Through counterfactual modelling and tracking the evolution in \emph{what} is discussed, \emph{who} is driving the discussions, and \emph{where} discussions take place, we offer a lens that we can unravel the mechanisms underlying opinion shift. 

We illustrate the utility of our approach by applying it to stance change towards UBI on Reddit using a combination of embedding techniques, cohort-level analysis, and a fine-tuned topic classifier. 
We find that macro stance shifts are in fact driven by micro variations in stance --- change within different topics of discussion, different user cohorts, and different communities of discussion. In each of these cases, how the stance changed is more important than shifts in the proportion of discussion.
Even though stance is more important, proportions for the partisan dimension and across user cohorts still play a statistically significant role in determining the overall stance of UBI discussions over time. 
The best counterfactual explanation for overall stance change is cohort stance change over time.

Recently, calls for UBI have been increasing. OpenAI is preparing for the prospect of Artificial General Intelligence (AGI) by rolling out a UBI program.\footnote{\url{https://openai.com/blog/planning-for-agi-and-beyond}} The state of Georgia has also piloted its own tests with UBI. With this reality setting in, it is critical to understand how different stakeholders and groups view, and will be affected by, UBI. The analyses in this paper provide several findings that will hopefully form future policy. First, since 2020, we are noticing an unprecedented concentration of positive stance towards UBI across cohorts. Prior to 2019, the difference in stance between new and old cohorts was over 0.13 points. Following 2020, this shift drops to less than 0.10 points. 
Second, we find an increase in polarization surrounding UBI following 2020. In particular, the difference in support across the extremes in the affluence dimension has grown. 
Finally, we find evidence that stance towards UBI and its subtopics is highly variable. What is one year a positive frame for discussing UBI, can quickly change to become a more negative frame (e.g., Labor wages and work conditions). This effect is more important for driving overall stance change, than shifts in proportion of subtopics being discussed. 

Furthermore, the only social dimension where polarization appears to grow between communities is affluence. The less affluent communities experience a disproportionate upswing in support for UBI, whereas more affluent communities do not change. Beyond this, we find little evidence of polarizing occurring in the partisan dimension. 

There are limitations to our research. It is important to note that the characteristics of Reddit influence our findings. For instance, Reddit users are not evenly distributed around the world. The platform caters mainly to a younger, English-speaking population; the top four nations representing the greatest share of Reddit traffic are the United States, United Kingdom, Canada, and Australia~\cite{clement2021reddit}. Additionally, using more advanced models may have improved entity linking, allowing for a higher recall of UBI-related comments than the keyword search. Some of the methods in our approach, such as the manual evaluation of topic coherence, require domain expertise. Moreover, since our topic model inferred topics from the entire dataset, it could have missed smaller topics that are pertinent in specific time periods. One direction for further research would be to perform the content-level analyses by inferring topics for different time periods separately.

\xhdr{Broader Perspective} With regards to data collection, we use publicly available Reddit data; all Reddit users agree to Reddit's Privacy Policy when they use the platform, which stipulates that a user's contributions can be shared publicly. 
While these comments may contain offensive content, our manual inspections of a random sample of comments shows that such comments do not occur often in the dataset. Furthermore, we only report aggregated results from our data which do not include any personally identifiable information.
Additionally, we plan to share our code upon publication.

With regards to the broader implications of our work, a potential negative use case would be applying our computational approach to understand different factions of a contentious topic. This knowledge could be input into a misinformation campaign aiming to construct comments with different framings. These framings could then be used in different communities and online social contexts to drive controversy and sow discord. 

However, we believe that our work poses few risks and has the potential to benefit many groups of people. We live in a world where more and more political discussions and debates are happening online. Platforms like Reddit are not only used by the general population, but are also used by politicians and political groups to communicate directly with the public. Our approach could help better understand the drivers behind overall stance chance on important issues. A clearer understanding of these drivers can help political organizations communicate more effectively with individuals in key demographics and draw the appropriate conclusions from platform-level shifts. In particular, a strength of our method is that it uncovers the many different groups discussing the same topic and helps quantify the sometimes-subtle distinctions in their perspectives over time. Compared with more expensive methods that are limited to less comprehensive analyses, this could be particularly helpful for minority groups.

\section{Acknowledgements}
We thank the Laidlaw Foundation, NSERC, CFI, and ORF for their support.

\bibliography{aaai25, custom}

\appendix
\section{Ethics Checklist}
\label{sec:ethics}

\begin{enumerate}

\item For most authors...
\begin{enumerate}
    \item  Would answering this research question advance science without violating social contracts, such as violating privacy norms, perpetuating unfair profiling, exacerbating the socio-economic divide, or implying disrespect to societies or cultures?
    \answerYes{Yes, and our findings are focused on understanding diverse topics, users, and communities contribute to overall discussions on an important political issue.}
  \item Do your main claims in the abstract and introduction accurately reflect the paper's contributions and scope?
    \answerYes{Yes, and the main contributions are first, a methodology to identify the fine-grained social drivers of opinion change on an online platform, and second, an application of this methodology on an important policy issue.}
   \item Do you clarify how the proposed methodological approach is appropriate for the claims made? 
    \answerYes{Yes, and we clarify why we decide to use the counterfactual setup in Section~\ref{sec:theory}}
   \item Do you clarify what are possible artifacts in the data used, given population-specific distributions?
    \answerYes{Yes, we specify the artifacts of Reddit data.}
  \item Did you describe the limitations of your work?
    \answerYes{Yes, and we discuss the limitations of Reddit data and our keyword-based method of gathering UBI comments.}
  \item Did you discuss any potential negative societal impacts of your work?
    \answerYes{Yes, and we believe that overall, our work poses few risks and has the potential to benefit many people.}
      \item Did you discuss any potential misuse of your work?
    \answerYes{Yes, and we believe that overall, our work poses few risks and has the potential to benefit many people.}
    \item Did you describe steps taken to prevent or mitigate potential negative outcomes of the research, such as data and model documentation, data anonymization, responsible release, access control, and the reproducibility of findings?
    \answerYes{Yes, and we have described our methodology in detail in the Appendix, and plan on creating a public repository with the code used for this project.}
  \item Have you read the ethics review guidelines and ensured that your paper conforms to them?
    \answerYes{Yes.}
\end{enumerate}

\item Additionally, if your study involves hypotheses testing...
\begin{enumerate}
  \item Did you clearly state the assumptions underlying all theoretical results?
    \answerNA{N/A}
  \item Have you provided justifications for all theoretical results?
    \answerNA{N/A}
  \item Did you discuss competing hypotheses or theories that might challenge or complement your theoretical results?
    \answerNA{N/A}
  \item Have you considered alternative mechanisms or explanations that might account for the same outcomes observed in your study?
    \answerNA{N/A}
  \item Did you address potential biases or limitations in your theoretical framework?
    \answerYes{Yes, and we highlight how our counterfactual analyses differ between three meta-factors we explore (user-level analysis v.s. content- or community-level analysis.}
  \item Have you related your theoretical results to the existing literature in social science?
    \answerNA{N/A}
  \item Did you discuss the implications of your theoretical results for policy, practice, or further research in the social science domain?
    \answerNA{N/A}
\end{enumerate}

\item Additionally, if you are including theoretical proofs...
\begin{enumerate}
  \item Did you state the full set of assumptions of all theoretical results?
    \answerNA{N/A}
	\item Did you include complete proofs of all theoretical results?
    \answerNA{N/A}
\end{enumerate}

\item Additionally, if you ran machine learning experiments...
\begin{enumerate}
  \item Did you include the code, data, and instructions needed to reproduce the main experimental results (either in the supplemental material or as a URL)?
    \answerNo{No, but we plan on making a public GitHub repository with this information upon publication}
    \item Did you specify all the training details (e.g., data splits, hyperparameters, how they were chosen)?
    \answerYes{Yes.}
    \item Did you report error bars (e.g., with respect to the random seed after running experiments multiple times)?
    \answerNo{No}
    \item Did you include the total amount of compute and the type of resources used (e.g., type of GPUs, internal cluster, or cloud provider)?
    \answerNo{No.}
     \item Do you justify how the proposed evaluation is sufficient and appropriate to the claims made? 
    \answerYes{Yes}
     \item Do you discuss what is ``the cost`` of misclassification and fault (in)tolerance?
    \answerNo{No}
  
\end{enumerate}

\item Additionally, if you are using existing assets (e.g., code, data, models) or curating/releasing new assets, \textbf{without compromising anonymity}...
\begin{enumerate}
  \item If your work uses existing assets, did you cite the creators?
    \answerNo{No, because our dataset is created by Reddit users, and there are hundreds of thousands of them in our dataset. However, we do cite Pushshift, the platform responsible for curating the Reddit data.}
  \item Did you mention the license of the assets?
    \answerNA{N/A}
  \item Did you include any new assets in the supplemental material or as a URL?
    \answerNA{N/A}
  \item Did you discuss whether and how consent was obtained from people whose data you're using/curating?
    \answerYes{Yes, and consent was obtained through Reddit users' agreeing to Reddit's Privacy Policy when creating an account.}
  \item Did you discuss whether the data you are using/curating contains personally identifiable information or offensive content?
    \answerYes{Yes, and we clarify that since our results are aggregated, we do not expose any personally identifiable information in this paper.}
\item If you are curating or releasing new datasets, did you discuss how you intend to make your datasets FAIR (see \citet{fair})?
\answerNA{N/A}
\item If you are curating or releasing new datasets, did you create a Datasheet for the Dataset (see \citet{gebru2021datasheets})? 
\answerNA{N/A}
\end{enumerate}

\item Additionally, if you used crowdsourcing or conducted research with human subjects, \textbf{without compromising anonymity}...
\begin{enumerate}
  \item Did you include the full text of instructions given to participants and screenshots?
    \answerNA{N/A}
  \item Did you describe any potential participant risks, with mentions of Institutional Review Board (IRB) approvals?
    \answerNA{N/A}
  \item Did you include the estimated hourly wage paid to participants and the total amount spent on participant compensation?
    \answerNA{N/A}
   \item Did you discuss how data is stored, shared, and deidentified?
   \answerNA{N/A}
\end{enumerate}

\end{enumerate}

\section{Appendix}
\label{sec:appendix}

\renewcommand{\thefigure}{A\arabic{figure}}
\setcounter{figure}{0}

\subsection{Additional User-Level Analyses}
 \begin{figure*}[h!]
    \centering
    \includegraphics[width=14.8cm]{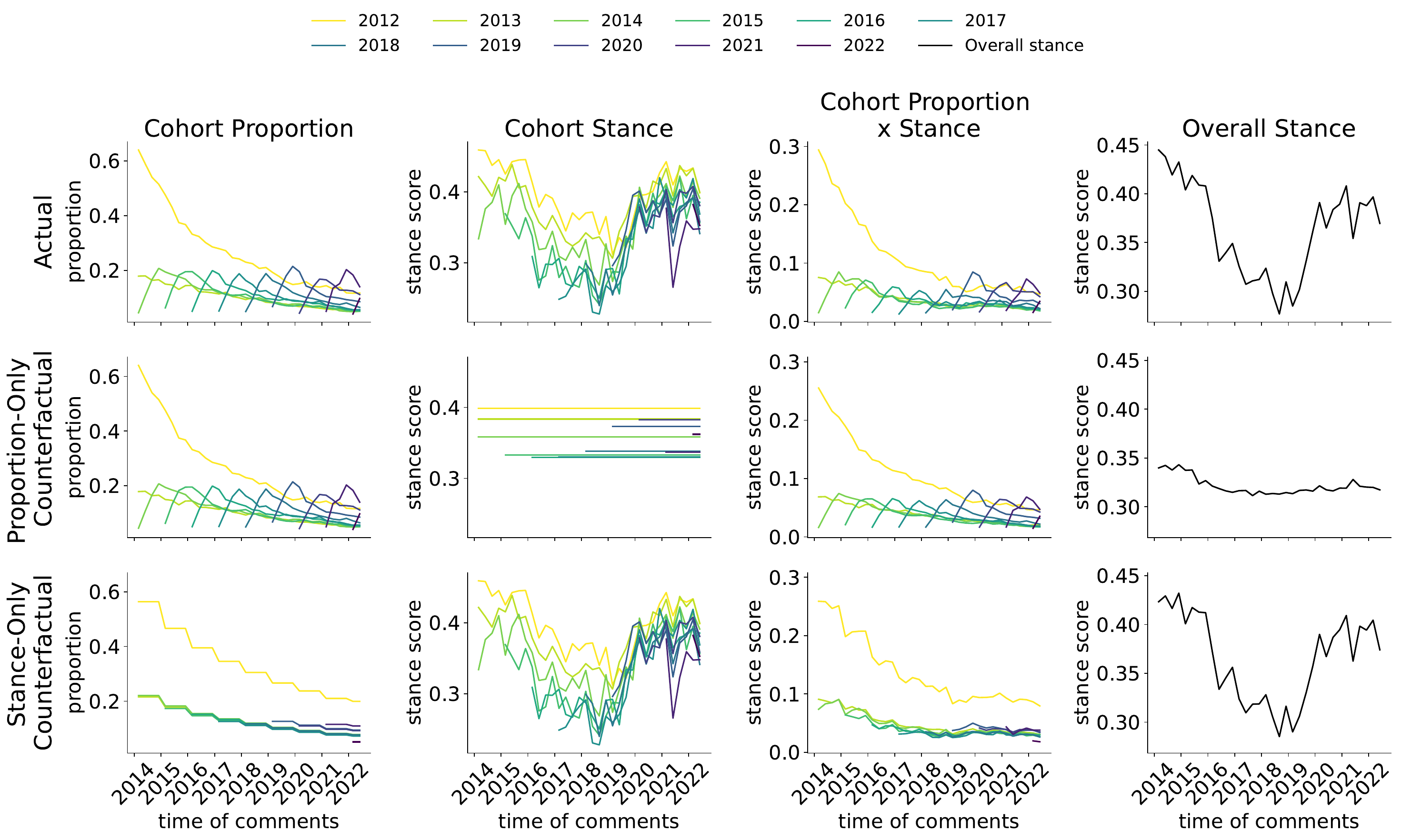}
    \caption{Actual, proportion-only counterfactual, and stance-only counterfactual scenarios. The top row illustrates how the proportions and stances in each cohort and the overall stance changed over time. The middle row illustrates the counterfactual scenario where we keep the stances constant, and let the proportions vary. The bottom row illustrates the counterfactual scenario where we keep the proportions constant, and let the stances vary. Note that for the bottom row, we renormalize the proportions for each cohort in each quarter so that the proportions add to 1.}
    \label{fig:appendix-cohort-3-by-4}
\end{figure*}

\subsection{Additional Community-Level Analyses}
\begin{figure}[h]
    \centering
    \includegraphics[width=7.7cm]{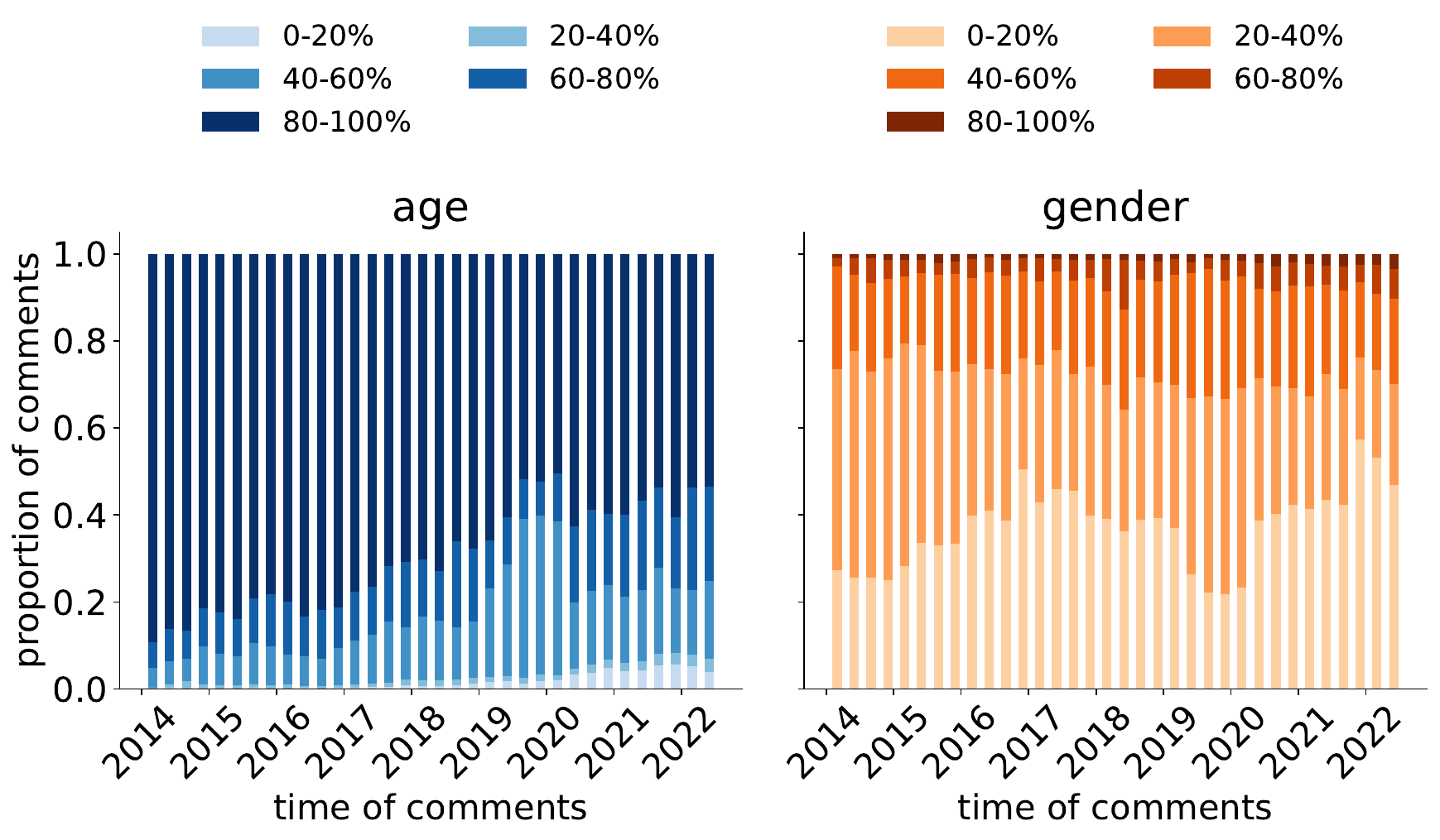}
    \caption{Proportion of UBI comments made by communities along the age and gender dimensions. Lighter colours represent younger or masculine communities, and darker colours represent older or feminine communities.}
    \label{fig:appendix-age-gender-proportion}
\end{figure}

\begin{figure}
    \centering
    \includegraphics[width=7.7cm]{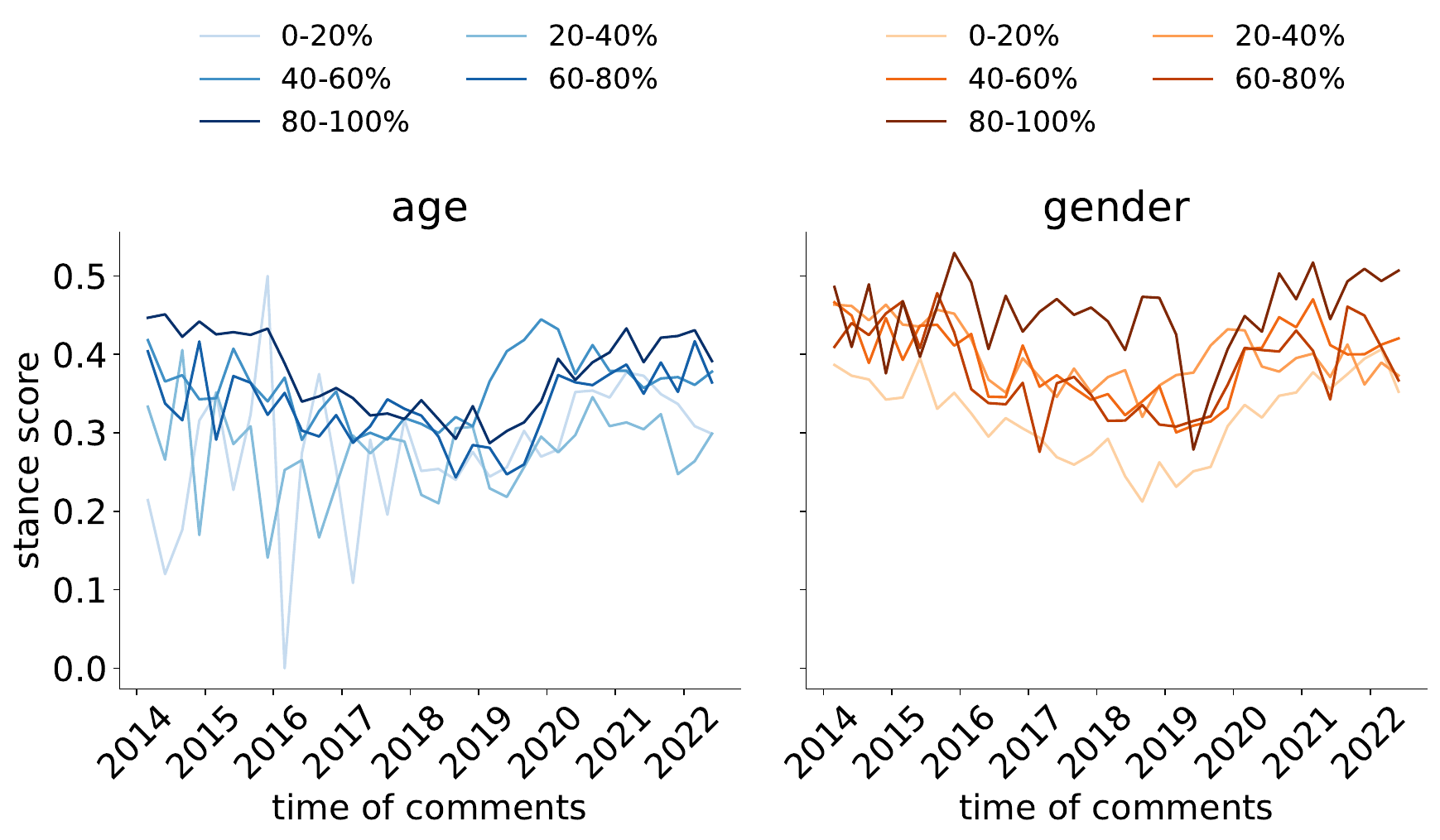}
    \caption{Average stance on UBI comments along the age and gender dimensions. Lighter colours represent younger or masculine communities, and darker colours represent older or feminine communities.}
    \label{fig:appendix-age-gender-stance}
\end{figure}

\begin{figure*}
    \centering
    \includegraphics[width=14.8cm]{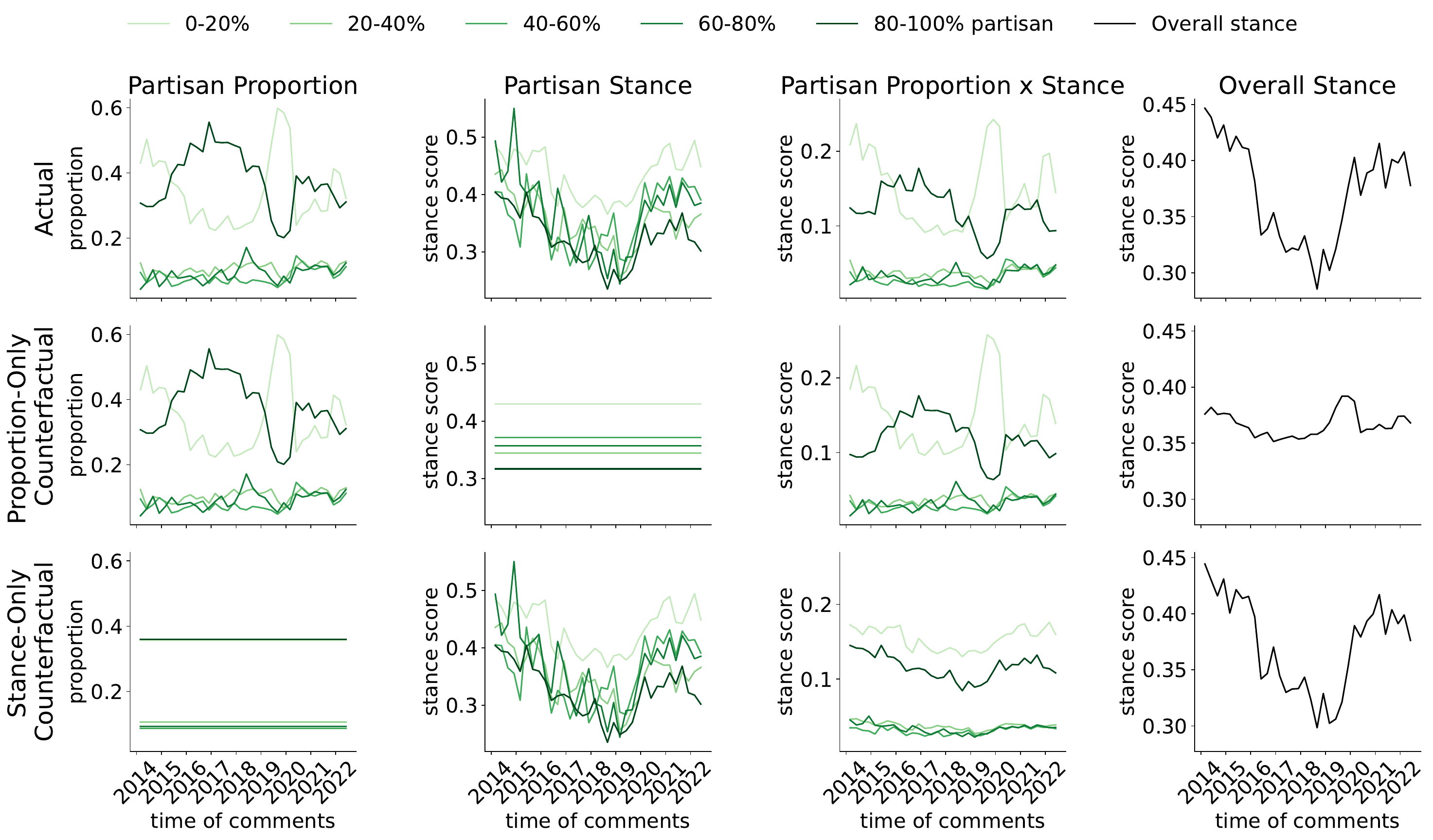}
    \caption{Actual, proportion-only counterfactual, and stance-only counterfactual scenarios. The top row illustrates how the proportions and stances in each partisan bin and the overall stance changed over time. The middle row illustrates the counterfactual scenario where we keep the stances constant, and let the proportions vary. The bottom row illustrates the counterfactual scenario where we keep the proportions constant, and let the stances vary.}
    \label{fig:appendix-partisan-3-by-4}
\end{figure*}

\begin{table}[h]
    \centering
    \begin{tabular}{l|c|c}
    \toprule
        Varying Property & Pearson $r$ & Loss \\
        \midrule
        Age Proportion & -0.111 & 1.351 \\
        Age Stance & 0.981 & 0.271 \\
        Gender Proportion & 0.312 & 1.286 \\
        Gender Stance & 0.988 & 0.198 \\
    \bottomrule
    \end{tabular}
    \caption{The Pearson $r$ and losses between each the age and gender counterfactual worlds and the true stance.}
    \label{tab:age-gender-counterfactuals}
\end{table}

\subsection{Data Preprocessing}\label{appendix:data}
\textbf{Exploration of more permissive approaches.} First, we considered taking all the discussion trees of a submission or comment with the keyword; however, upon analyzing sample submissions, such as ``Pope Calls for Universal Basic Income, Shorter Working Day,'' we found that discussions deviated from the topic of UBI quite quickly; for instance, there were children of top-level comments that were solely about bishops' retirement. Second, we considered a ``snowball sampling'' approach, similar to the one used by Su et al. \shortcite{SU20191}. We found the top 100 words that had the highest tf-idf scores within the dataset collected using the previously-mentioned keywords, but found that the words (ex. ``people,'' ``income,'' ``work'') were too general.

We also chose to focus on comments rather than submissions since comments usually contain more textual content and are more likely to occur in a wide-range of contexts, whereas submissions tend to center around a few communities and often link to external websites (such as news articles). These observations indicated that it was more likely for comments to contain organic \emph{discussions} surrounding UBI, which was what we were most interested in capturing.

\textbf{Additional filtering.} To avoid comments that use ``UBI'' in the middle of a word, we filter for comments that contain ``UBI'' preceded and followed by a non-alphanumeric character, ``UBI'' at the beginning of a comment followed by a non-alphanumeric character, or ``UBI'' at the end of a comment preceded by a non-alphanumeric character. We also remove any comments containing the phrase ``Ubisoft'' (case insensitive). Reddit is comprised of many different communities, called ``subreddits,'' each dedicated to discussing a specific topic; each comment on Reddit is associated with exactly one subreddit. Upon looking at the top 100 subreddits with the most UBI comments, which account for 71.5\% of all the UBI comments in the original dataset, a couple of subreddits seem irrelevant to UBI, such as r/Rainbow6, a subreddit dedicated to a video game. Many of these gaming subreddits use the phrase ``UBI'' to refer to Ubisoft. We identify six subreddits (r/Rainbow6, r/forhonor, r/thedivision, r/GhostRecon, r/Thread\_crawler, r/assassinscreed) from the top 100 subreddits that were unlikely to use ``UBI'' to refer to universal basic income. To reduce irrelevant comments, we filter out any comments containing just ``UBI'' and not ``basic income'' from these communities since the phrase ``basic income'' is more specific. We remove comments from the bots u/AutoModerator, u/assessment\_bot, u/subredditsummarybot, u/transcribot, u/SnapshillBot, u/sneakpeekbot, u/twitterInfo\_bot, and u/autowikibot.
Finally, we take comments from January 2014 onward since there was very little UBI discussion before this date. 

\subsection{Stance Detection}\label{appendix:stance}
\xhdr{GPT-4 prompting strategy} We used the following prompt to label each comment. The prompt takes the text of the comment we want to label, and the text of the parent comment or submission.

\begin{lstlisting}[language={}]
    You are a model that answers the question "what is the attitude of the commenter towards UBI?" based on the Reddit comment they made.
    
    Each commenter's stance can be categorized under one of the following three stances. 
    - "Neutral"
    - "Against"
    - "Supportive"  
    
    If the commenter does not explicitly provide their own personal opinion on UBI, their stance should be classified as "Neutral". Merely stating a fact that would imply opposition towards UBI is not enough to classify the commenter's stance as "Against". In this case, the commenter's stance should be assigned as "Neutral". Merely stating a fact that would imply support for UBI is not enough to classify the commenter's stance as "Supportive". In this case, the commenter's stance should be assigned as "Neutral". 
    
    You should only respond with one of the three phrases above.
    
    Here is the comment. Identify the commenter's stance on UBI based on this comment:
    {}
    
    For context, the above comment was a response to the following comment. Do not identify the stance of the author of this comment:
    {}
\end{lstlisting}

\xhdr{Stance categories} A commenter's attitude towards UBI can either be supportive (indicating that they support the implementation of the policy), against (indicating that they are against the implementation of UBI), or neutral, if the comment is a statement of fact, or a question, or if the commenter does not explicitly provide their own personal opinion on UBI.

Some comments could be interpreted as both supportive / neutral or neutral / against. An example is the following comment: 

\begin{lstlisting}[language={}]
The way you can figure out how long your job will last is to pay attention to what you spend most of your time doing. If you feel like your job is mostly just going through the motions where you look at something then make decisions based on preset criteria. \n\nPolish your resume and beg for Universal Basic Income as soon as possible. Your job is going to go away.\n\nIf you spend your time building things and solving complex problems. You will last longer.
\end{lstlisting}

This comment was labelled by three different people. Two people labelled the comment as neutral and one person labelled it as supportive. 
Despite this ambiguity in some comments, when testing our classifier, we labelled each comment with only one stance.

\subsection{Topic Modelling Details}\label{appendix:topics}
\xhdr{Preprocessing steps} 
We first perform a sequence of preprocessing steps.
First, we use the Python langdetect library to detect the language of the dataset using the first 15 words of each comment. We found 16,275 non-English comments. Since most of the comments in our dataset are in English, we translate all of the non-English comments so that the topic model does not group all the non-English comments into one topic. 
Second, we convert all text into lowercase and remove all URLs. 
Third, we remove the phrases ``universal basic income'', ``unconditional basic income'', ``basic income'', and ``ubi'' from the comments, since we are interested in what subtopics lie \emph{within} UBI discussion and do not want UBI to be a subtopic itself. We also remove all other stopwords in the nltk library, non-alphanumeric characters, and words shorter than 3 letters or longer than 15 letters. 
Sixth, we lemmatize the text. From the lemmatized text, we again filter out words shorter than 3 characters. 
Seventh, we compute all bigrams in the text using the gensim library, taking min\_count $=$ 150 and threshold $=$ 25.
Eighth, we filter out words used in fewer than 150 documents and words used in more than 50\% of the documents.

\xhdr{Model parameters} We use the gensim library for LDA. We keep most of the default parameters provided by gensim for the model, and set passes to 5 and the hyperparameters alpha to `auto' and eta to `auto' to learn these values from the data.

To find the ideal number of topics, we experiment with various numbers of topics and determine the ``goodness'' of a certain number of topics using the topic coherence measure $C_V$. The possible values of $C_V$ range from 0 to 1, with 1 denoting perfect coherence and 0 representing complete incoherence. We choose $C_V$ over other topic coherence measures since it correlates most with human judgements on the coherence of topics \cite{roder_2015}.

To determine roughly the ideal number of topics, we first create LDA models with 5 to 95 topics, at increments of 10 topics. The topic coherence is highest at 15 topics and declines overall for higher values. To converge on an exact number of topics, we create LDA models with 5 to 25 topics. We find that setting the number of topics to 19 produces the model with the highest coherence ($0.53$).

We manually validate the quality of the topics produced as follows. 
First, we name the topics based on the top 10 keywords.
Second, borrowing from the manual validation methodology used by Aslett et al. \shortcite{aslett2022parkland}, we choose a ``threshold'' probability of $0.2$. Aslett et al. \shortcite{aslett2022parkland} use the probability threshold of $0.1$ in their work; we choose a higher value since we have less topics (19 v.s. 32). For each topic, we randomly sample 50 comments from the comments that pass this threshold probability for the topic, and determine whether they belong to the particular topic.
For most topics, we found that at least 85\% of the comments belong to the topic.
In some cases, we find that some topic names can be broadened (within reason). For instance, we originally named a topic as ``Wages'' based on the top 10 keywords, but found that some comments from our random sample also mention a reduction in working hours. So, we rename the topic ``Conditions for Work.''
In other cases, we find that some topics cannot be broadened. For instance, we originally named a topic as ``Common Words,'' but found that the comments included discussions on automation, welfare, and human nature, which indicate that the topic is too broad to be coherent. So, we discarded this topic.
We also discarded topics that were too narrow, such as topics containing a significant number of comments made by bots.
All together, we discarded 4 topics, leaving us with a final topic model with 15 topics, which we highlight in Table \ref{tab:topics_overview}.

\begin{table*}[t!]
    \tiny
    \centering
    \begin{tabular}{l|l}
    \toprule
        Subtopic Name & Top 10 Keywords \\
        \midrule
        Living costs & price, rent, housing, cost, increase, land, landlord, demand, market, supply \\
        Data analysis and research & study, read, book, show, data, effect, article, research, experiment, link  \\ 
        Education and family & child, kid, school, parent, family, woman, young, college, sex, student  \\
        Non-UBI government welfare programs & would, government, benefit, program, welfare, system, social, cost, current, cut  \\
        Budget and cost & year, month, per, million, dollar, every, billion, average, budget, trillion \\
        Economic systems & system, society, capitalism, value, market, economic, economy, power, human, create \\
        Labor wages and work conditions & job, work, wage, worker, minimum\_wage, pay, hour, low, give, everyone \\
        Public services and healthcare & healthcare, free, public, education, universal\_healthcare, housing, law, drug, health\_care, medical \\
        Money and inflation & money, inflation, debt, economy, bank, currency, loan, print, growth, cause \\
        Politics and elections & vote, party, politician, policy, political, win, election, politics, candidate, green \\
        Global affairs & country, american, world, america, nation, war, global, china, usa, citizen  \\
        Automation and jobs & automation, job, automate, robot, human, technology, machine, replace, future, industry \\
        Taxes & tax, income, pay, fund, rate, high, increase, low, rich, wealth \\
        Political affiliations & conservative, anti, support, right, libertarian, liberal, left, pro, democrat, progressive \\
        Businesses and profit & business, company, profit, corporation, owner, market, worker, stock, union, product \\
    \bottomrule
    \end{tabular}
    \caption{Subtopic name and top 10 keywords produced by LDA.}
    \label{tab:topics_overview}
\end{table*}

To serve as a baseline, we calculate the maximum macro-F1 and micro-F1 we can get using the LDA model on the random sample of 95 comments used to validate our stance and topic models. LDA provides a distribution over all the topics for each comment. To evaluate the LDA model, we experiment with different ``threshold'' probabilities: a comment is classified into all the topics in its topic distribution that exceed this probability. We choose two threshold probabilities: one that maximizes the macro-F1 (0.11) and another that maximizes the micro-F1 (0.24). We find that the best macro-F1 we get is 0.393 and the best micro-F1 is 0.417, which is lower than what we get using our topic model.

\xhdr{\turbo{} prompting strategy} We used the following prompt to label each comment. The prompt takes the top four LDA topics, a short description for each of them (which was created by reading the top 10 LDA words and comments and extracting the common themes), the text of the comment we want to label, and the text of the parent comment or submission. The descriptions of each topic are in~\ref{tab:topics_descriptions} and the prompt is below:

\begin{lstlisting}[language={}]
    You are a model that classifies a UBI-related Reddit comment into a taxonomy. 

    Taxonomy:
    - {}. {} 
    - {}. {}  
    - {}. {}  
    - {}. {}  
    - None of the above

    You should only respond in a Python list format. Make the list as short or as long as you need. Each element in the list should be a class in the taxonomy. Do not add anything after the Python list.

    Here is the comment. Classify this comment: 
    {}.

    For context, the above comment was a response to the following comment. Do not classify this comment:
    {}.
\end{lstlisting}

\subsection{Community Embeddings}\label{appendix:community}
Community embeddings position subreddits in a 150-dimensional space using the word2vec algorithm, treating communities as ``words'' and commenters as ``contexts''. In the word2vec word embedding, the vectors of words used in similar contexts are closer together; analogously, the vectors of communities with similar users are closer together in the community embedding. In the community embedding, Waller and Anderson~\shortcite{waller2021quantifying} represent the largest 10,006 communities by the number of comments from 2005 to the end of 2018 by 150-dimensional vectors, accounting for 95.4\% of all comments on Reddit in this timeframe. Notably, the closeness of any two communities within this embedding is entirely contingent on \textit{who} comments in the communities and not \textit{what} they comment. 

We apply the exact methodology used by Waller and Anderson but extend the timeline to June 2022. Since the community embeddings include only a subset of the all Reddit communities, we take only the comments from our dataset that belong to a subreddit represented in the community embeddings for all our analyses involving the social dimensions. There were a total of $1,115,660$ comments that satisfied this criterion.

Using the community embeddings, we can define social dimensions. These social dimensions enable us to position any Reddit community across any social dimensions of interest; for instance, we can quantify how left- or right-leaning a community is.
Social dimensions are found from the community embeddings by first seeding a pair of communities that differ in the social dimension of interest but are similar in other regards. These social dimensions provide further insight into the communities on Reddit, for instance, whether a community is more feminine or masculine. We consider the following relevant social dimensions: age (seeded with r/teenagers and r/RedditForGrownups), gender (seeded with r/AskMen and r/AskWomen), partisan (seeded with r/democrats and r/Conservative), and affluence (seeded with r/vagabond and r/backpacking) \cite{waller2021quantifying}. The initial social dimensions are created by taking the vector difference between the two seeds in each seed pair. Waller and Anderson \shortcite{waller2021quantifying} include additional steps such as the augmentation of seed pairs using nine other similar pairs of communities to make the social dimensions more robust. This process gives us a vector for each social dimension in the community embedding space. To get the social scores for a subreddit (ex. partisan-leaning), the vector representing the Reddit community is projected onto the vector representing the relevant social dimension (ex. the partisan-leaning vector). Then, the projections are assigned a percentile score depending on where the community is positioned along the relevant social dimension. For instance, for the partisan dimension, the most left-wing community would be assigned a score of 0, while the most right-wing community is assigned a score of 100. In the social dimensions, a higher social score for a community indicates that a community leans older, more feminine, more right-wing, and more affluent for the age, gender, partisan, and affluence dimensions, respectively.

\subsection{Additional Time Series Similarity Measures}

We include the results from two other time series similarity metrics---Euclidean distance and Dynamic Time Warping (DTW). For both metrics, we find that the same overall result holds: stance changes within the topics, users, and communities of discussion are more responsible for overall stance change than distribution shifts.

\begin{table}[]
    \centering
    \begin{tabular}{c|c|c}
         Counterfactual & Euclidean Distance & DTW \\
         \hline
         Subtopic Proportion & 0.266 & 0.203 \\ 
         Subtopic Stance & 0.073 & 0.056 \\
         Cohort Proportion & 0.336 & 0.310 \\
         Cohort Stance & 0.037 & 0.036 \\
         Partisan Proportion & 0.234 & 0.189 \\
         Partisan Stance & 0.062 & 0.048 \\
         Affluence Proportion & 0.246 & 0.214 \\
         Affluence Stance & 0.051 & 0.048 \\
         Age Proportion & 0.258 & 0.250 \\
         Age Stance & 0.060 & 0.047 \\
         Gender Proportion & 0.246 & 0.216 \\
         Gender Stance & 0.044 & 0.038 \\
    \end{tabular}
    \caption{The Euclidean distance and DTW distance between each counterfactual and the true stance. The Euclidean distance is calculated by considering each time series as a vector. A higher distance indicates that the counterfactual is less similar to the true stance.}
    \label{tab:additional_similarity_measures}
\end{table}

\end{document}